\documentclass[%
 sc,
 amsmath,amssymb,			
 reprint,
 nobalancelastpage,%
]{revtex4-1}
\usepackage[utf8]{inputenc}
\usepackage{graphicx}
\usepackage{dcolumn}
\usepackage{bm}
\usepackage{hyperref}
\usepackage{amsmath}
\usepackage{subcaption}
\hypersetup{colorlinks,colorlinks,linkcolor=cyan,anchorcolor=cyan,citecolor=cyan,filecolor=cyan,urlcolor=cyan}
\usepackage[utf8]{inputenc}
\usepackage[T1]{fontenc}
\usepackage{mathptmx}
\usepackage{etoolbox}
\usepackage{lmodern}
\usepackage[capitalize]{cleveref}

\RequirePackage{xpatch}
\xpatchcmd\citenum{\NAT@parfalse}{\NAT@partrue}{}{}

\makeatletter
\def\@email#1#2{%
 \endgroup		
 \patchcmd{\titleblock@produce}
  {\frontmatter@RRAPformat}
  {\frontmatter@RRAPformat{\produce@RRAP{*#1\href{mailto:#2}{#2}}}\frontmatter@RRAPformat}
  {}{}
}%
\makeatother
\begin{document}

\preprint{AIP/123-QED}

\title{A Novel Linked-Mirror Device for Fusion}

\author{Haorong Qiu$^1$}\author{Xiang Li$^1$}\author{Guodong Yu$^2$}\author{Zhexuan Zheng$^2$}\author{Caoxiang Zhu$^2$}\author{Guoyong Fu$^{1,3}$}
\thanks{corresponding author's Email: gyfu@zju.edu.cn}

\affiliation{$^1$Institute for Fusion Theory and Simulation, School of Physics, Zhejiang University, Hangzhou, China}
\affiliation{$^2$School of Nuclear Science and Technology, University of Science and Technology of China, Hefei, China}
\affiliation{$^3$Jingjie Fusion (Hangzhou) Technology Co., Ltd, Hangzhou, China}

\date{\today}

\begin{abstract}
We design a novel linked-mirror device (LMD) with a magnetic well using the near-axis expansion (NAE) method. The design consists of two parallel straight mirror sections connected at both ends by three-dimensional stellarator-like magnetic fields , thereby eliminating the end-loss problem of a single magnetic mirror. By designing appropriate three-dimensional magnetic fields, we obtain an LMD configuration with a magnetic well, ensuring stability against interchange modes. The existence of finite-$\beta$ equilibria of the LMD is demonstrated by numerical solutions of the 3D equilibrium code GVEC. This design is potentially a novel solution to magnetic confinement for fusion reactors.\end{abstract}
\maketitle	
\section{Introduction}

Since the early 1950s, the fusion community has devoted sustained effort to developing magnetically controlled fusion energy, 
a field that holds immense promise as an abundant energy source for humanity. 
Over the past seven decades, a wide variety of magnetic confinement devices have been conceived and developed to harness fusion energy, including magnetic mirrors\cite{post1987magnetic}, stellarators\cite{spitzer1958stellarator}, and tokamaks\cite{azizov2012tokamaks}. 
Among these, the magnetic mirror stands out as one of the simplest magnetic confinement devices, 
serving as an important starting point in the exploration of magnetic fusion technology. 

In the initial stage of its development, however, magnetic mirror devices were plagued by two long-standing problems that hindered their practical application to fusion energy. 
First, there is the problem of rapid end losses of particles from both ends of a mirror machine\cite{baldwin1977end}. 
Second, the traditional magnetic mirror is susceptible to magnetohydrodynamics (MHD) interchange instability\cite{rosenbluth1957stability}. 
Of these two challenges, the rapid end loss of particles proved particularly detrimental to plasma confinement, making it difficult to achieve the high-temperature, high-density plasma conditions required for fusion ignition. 

In response to these pressing issues, the research community has since explored a range of technical solutions aimed at mitigating the fast end losses. 
Representative approaches include the tandem mirror, which enhances confinement via end plugs that create positive electrostatic potential barriers\cite{post1987magnetic}, 
and the more recent gas dynamic trap (GDT)\cite{ivanov2017gas}, which uses an “ambipolar mirror” plug to reduce the end losses. 

Recently, a linear-torus hybrid device featuring the linked-mirror configuration (LMC)\cite{feng2021proposal} has been proposed. The core design concept of this new configuration draws inspiration from two well-established concepts: the linear magnetic mirror and the figure-eight stellarator. 
Structurally, this hybrid configuration is composed of two straight mirror segments, which are connected by a pair of half-torus segments. 
A distinctive structural feature of the LMC is that the two linear mirror segments are not parallel but instead are inclined at a finite angle to each other. This specific angular arrangement gives rise to a three-dimensional magnetic field topology which generates a finite rotational transform, just like the rotational transform of the early figure-eight stellarator. 
As a direct result of this unique magnetic field topology, toroidally passing particles are effectively confined and the end losses are eliminated. 
The simulation results showed that the single-particle confinement is favorable and is comparable to that of an equivalent tokamak\cite{feng2021proposal}. 
However, the LMC has a magnetic hill of approximately $1.5\%$, which destabilizes interchange modes.

To solve this problem, we have designed a new linked-mirror device with a magnetic well in the present work. In this new design, the two straight mirror sections are parallel and are linked by 3D magnetic fields at both ends. The 3D linking fields generate both a finite rotational transform and a finite magnetic well
while preserving the basic structure of the mirror sections, ensuring stability against interchange modes. The new LMD configuration was designed using the near-axis expansion (NAE) method\cite{garren1991magnetic}.
Subsequently, we have verified the results of the NAE solution for 
the vacuum magnetic field of LMD using the 3D MHD equilibrium code GVEC\cite{hindenlang2025computing}. For the first time, numerical equilibria of the linked-mirror configuration have been computed at finite beta using GVEC. Finally, we have designed a coil set with both circular coils and 3D coils using the FOCUS code\cite{zhu2018new}. The obtained coil set reproduces the vacuum magnetic field of the NAE solution, confirming the existence of a magnetic well. Particle simulations
 have also been performed using the FP3D code\cite{jiang2024fp3d}, which indicate good particle confinement for the LMD.

It should be noted that the linked-mirror approach has been considered previously. The earlier ELMO bumpy torus (EBT) consisted of many mirrors linked toroidally\cite{glowienka1981elmo}. Later, in 1981, the DRAKON closed fusion trap, proposed by Glagolev et al.~\cite{glagolev1981closed,kondakov2000numerical}, consisted of two long linear mirror sections joined by two curvilinear sections. More recently, Mazzucato proposed a fusion reactor scheme consisting of two linear mirrors joined by two toroidal sections with helical coils~\cite{mazzucato2020fusion}. Conceptually, our linked-mirror approach is similar to those of the DRAKON~\cite{kondakov2000numerical} and Mazzucato~\cite{mazzucato2020fusion}. The main difference is that our LMD design is based on analytic theory with rigorous numerical verification. Furthermore, our design possesses a favorable magnetic well that prevents the interchange instability of a linear mirror. Finally, a relatively simple coil set has been designed for the LMD, making it a novel solution for fusion reactors. 

The paper is organized as follows. 
Section II presents results of the LMD design, including the NAE equilibria and their numerical verification as well as the coil design. 
Section III discusses the results and future work. 
Section IV introduces the NAE method and the codes used in our design. 

\section{Results}
\subsection{Design of linked mirror using near-axis expansion method}
Here we present results of our new linked-mirror design obtained using the NAE method\cite{garren1991existence,garren1991magnetic}. The details of the NAE method are described in section IV.  Compared with the previous linked-mirror design, the two linear mirror sections are parallel to each other, and the rotational transform is generated by 3D magnetic fields connecting the two mirror sections. Specifically, the two half-torus sections have the shape of a rotating ellipse, as shown in Fig. 6. Furthermore, our design possesses a magnetic well arising from the triangular deformation of the elliptical magnetic surfaces.  
Below, we present the design in detail, including specification of the magnetic axis and the local coordinates used in the NAE method, the first-order NAE solution and the second-order NAE solution. The first-order solution specifies the rotating elliptical shape of the magnetic flux surface in the toroidal sections, and the second-order solution specifies the triangular deformation of the ellipse. 
It should be noted that we maintain a circular magnetic surface shape in the mirror sections (see Fig. 5). This is important for good confinement of trapped particles in the mirror sections. 

\subsubsection{Specification of the magnetic axis and local coordinates}

The NAE is performed by radially expanding the equilibrium magnetic field from the magnetic axis. Fig. 1 illustrates the magnetic axis and the corresponding coordinate system employed in our design. The magnetic axis consists of two parallel straight segments connected by two half-circles. The entire magnetic axis lies in a plane. The main parameters of the axis include the length of the mirror section $L_m=4.8\,\mathrm{m}$ and the radius of the half-circles $d=1.2\,\mathrm{m}$. The mirror ratio of the magnetic field is 4.

A local Frenet-Serret coordinate\cite{kabel2003maxwell} can be constructed based on this magnetic axis. 
First, we let L denote the total length of the magnetic-axis curve $\textbf{r}_{0}(s)$, and $0 \leq s \leq L$.
On the magnetic axis, a right-handed orthogonal unit vector system $( \textbf{t}, \textbf{n}, \textbf{b} )$ is defined. 
$\textbf{t}$ is the unit tangent vector along the magnetic axis, $\textbf{n}$ is the unit normal vector in the direction of curvature of the axis,  
$\textbf{b}$ is the unit binormal vector, which is perpendicular to both $\textbf{t}$ and $\textbf{n}$. 

\begin{figure}[!t]
	\centering
	\includegraphics[width=0.5\textwidth, keepaspectratio]{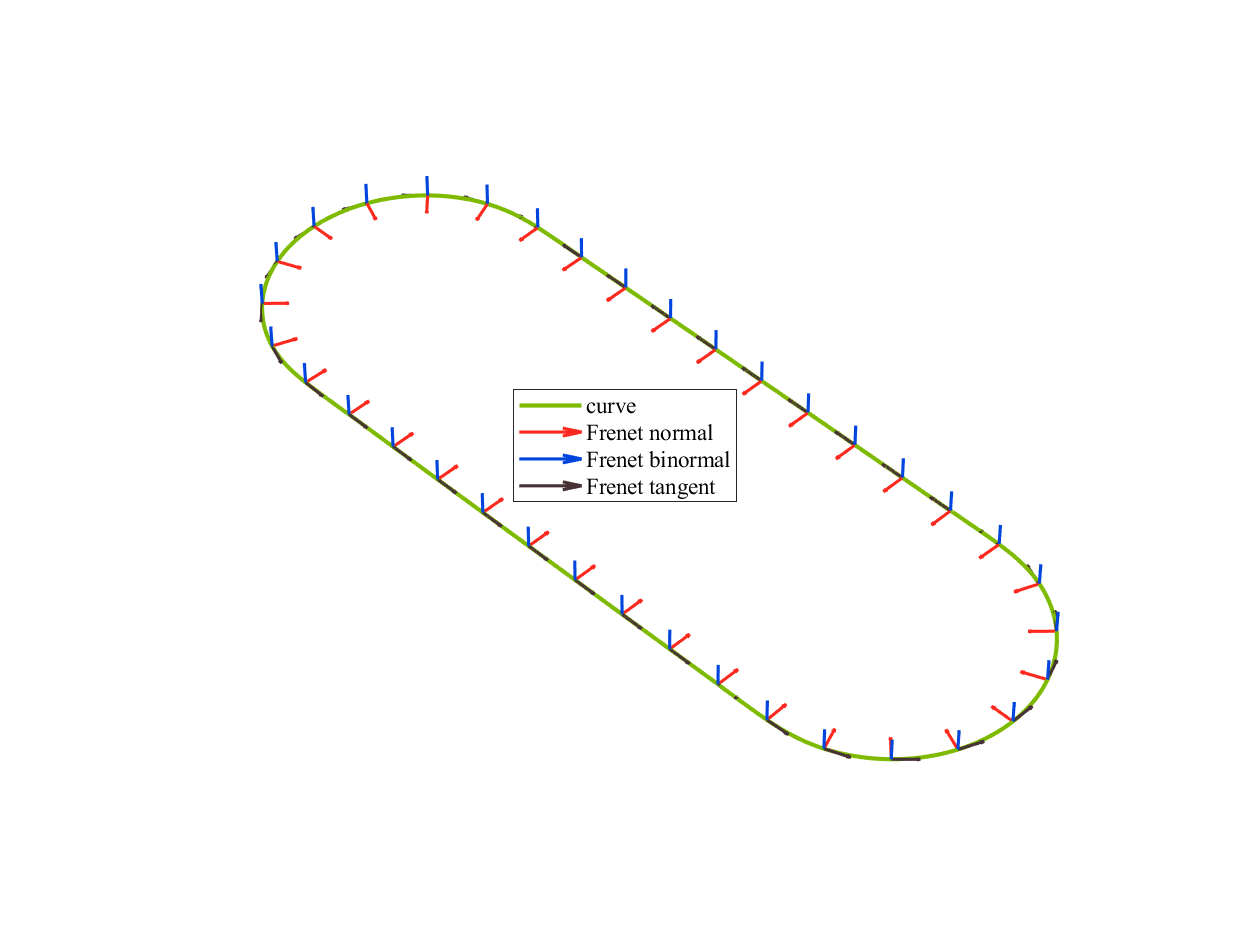} 	
	\caption{The magnetic axis and its Frenet-Serret frame}
	\label{axis}
\end{figure}

\subsubsection{The first-order NAE solution}

In an MHD equilibrium under Boozer coordinate system\cite{boozer1981plasma}, the coordinates of any point on the magnetic surface can be expressed as follows: 
\begin{align}
    \textbf{r}(r,\theta,\varphi) &= \textbf{r}_{0}(\varphi) + X(r,\theta,\varphi) \textbf{n}(\varphi) \nonumber \\
	&+ Y(r,\theta,\varphi) \textbf{b}(\varphi) + Z(r,\theta,\varphi) \textbf{t}(\varphi)
\end{align} 
where $\theta$ and $\varphi$ denote the Boozer poloidal and toroidal angles, $r(\psi)$ is an effective minor radius
defined by $2\pi\psi = \pi r^2 \overline{B}$, $\overline{B}$ is a constant reference field strength, $2\pi\psi$ 
is the toroidal flux. Here $\textbf{r}$, $\textbf{n}$, $\textbf{t}$ and $\textbf{b}$ are functions of $s$ with $s = s(\varphi)$ in the Boozer 
coordinates. Therefore, they are also functions of the toroidal angle alone. 

Let $R$ denote the scale length of the axis. If $r/R \ll 1$, the coefficients $X$, $Y$ and $Z$ can be expanded in the following form:
\begin{align}
    X(\psi,\theta,\varphi)&=r X_1(\theta,\varphi)+r^2 X_2(\theta,\varphi)+r^3 X_3(\theta,\varphi)+\cdots
\end{align}
The results of the expansion are substituted into the equilibrium equations. The NAE equilibria can be obtained order by order. For simplicity, our equilibrium design is performed under the zero-pressure condition. 

It can be seen that this is the parametric equation of an ellipse, 
and cross-sections at constant $\varphi$ are elliptical at first order in the expansion.

The first-order solution can be expressed in an intuitive form as follows\cite{garren1991magnetic}:
\begin{align}
X_1(\theta,\varphi) &= \Gamma(\varphi) \left[ cos(\theta - \gamma(\varphi)) + \xi(\varphi) cos(\theta - \gamma(\varphi) - \Delta(\varphi)) \right]\\
Y_1(\theta,\varphi) &= \Gamma(\varphi) \left[ sin(\theta - \gamma(\varphi)) - \xi(\varphi) sin(\theta - \gamma(\varphi) - \Delta(\varphi)) \right]
\end{align}
The shape of a surface at constant r can be described accordingly as follows: 
\begin{align}
	\textbf{r}^2(\theta,\varphi) &= X_1^2(\theta,\varphi) + Y_1^2(\theta,\varphi) \nonumber \\
    &= \Gamma^2(\varphi) \left[ 1 + \xi^2(\varphi) + 2\xi(\varphi) cos(2\theta - 2\gamma(\varphi) - \Delta(\varphi)) \right]
\end{align}
where $\Gamma^2(\varphi) = \frac{2}{B_0(\varphi)(1 - \xi^2(\varphi))}$, 
$\xi$ denotes a measure of the ellipticity, $\Delta$ denotes the angle between the ellipse and the X-axis, as shown in Fig. 2, $\Gamma(1+\xi^2)^{1/2}$ denotes the average minor radius of the ellipse, $B_0(\varphi)$ is the magnetic field strength on the magnetic axis. 
Fig. 2a shows the first-order magnetic surface shape and the shape parameters. The function $\gamma$ can be solved from the first-order equilibrium equations in terms of the other two variables. The functions $B_0(\varphi)$, $\xi(\varphi)$ and $\Delta(\varphi)$ can be freely prescribed. These functions, together with $\gamma(\varphi)$ and geometry parameters of the magnetic axis, completely specify the first-order NAE solution (vacuum equilibrium solution in this case). 

Given the above information, the rotational transform is given by the following equation:
\begin{align}
	\iota_0 &= \frac{1}{2\pi} \int_{0}^{2\pi} d\varphi \left[ \frac{ds}{d\varphi} (1-\xi^2(\varphi)) \tau(\varphi) 
	+ \frac{\xi^2(\varphi)}{1+\xi^2(\varphi)} \frac{d\Delta}{d\varphi} \right]
\end{align}
where $\tau(\varphi)$ is the torsion of the magnetic axis. This expression shows that the rotational transform arises from two terms. 
The first term is related to the torsion of the magnetic axis (i.e., the 3D structure of the magnetic axis) which is responsible for the generation of the rotational transform in the previous design of linked mirror. In the present design, this term does not contribute since the torsion vanishes. 
The second term corresponds to the rotation of the elliptical surface in the toroidal sections. In the present design, 
the rotational transform is entirely generated by this term. 
Furthermore, since the magnetic surface shape of the mirror segments is kept circular, this term does not contribute to the rotational transform. 

Fig. 3 shows that the rotational transform increases as $\xi_0$ increases. 
We choose the value of $\xi_0$ to be $\xi_0 = 0.41$. Correspondingly, the value of rotational transform is $\iota = 0.16$. 

\begin{figure}[!t]
	\centering
	\begin{subfigure}[b]{0.5\textwidth}
		\includegraphics[width=\textwidth]{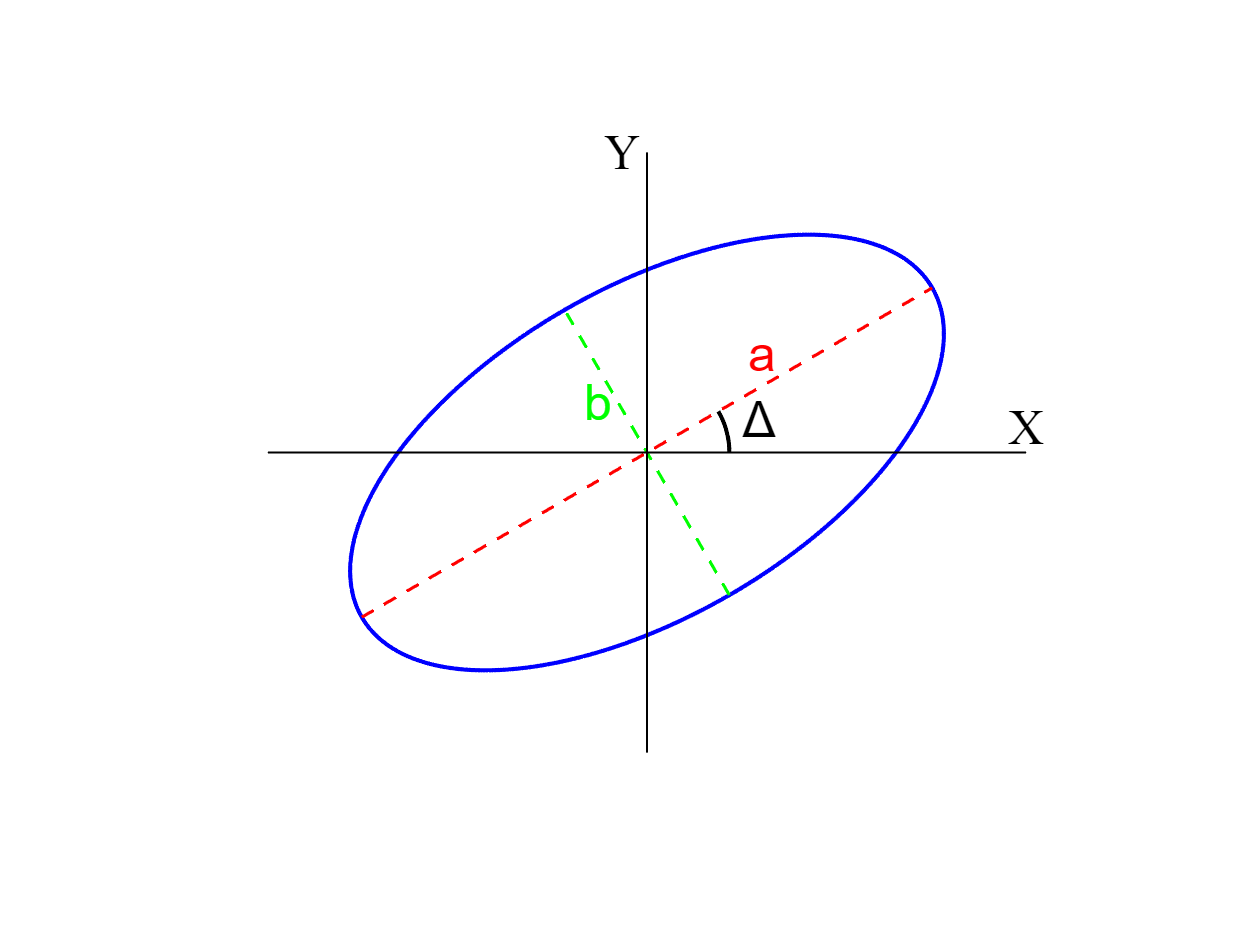} 
		\caption{}
		\label{ell.a}
	\end{subfigure}
	\quad
	\begin{subfigure}[b]{0.4\textwidth}
		\includegraphics[width=\textwidth]{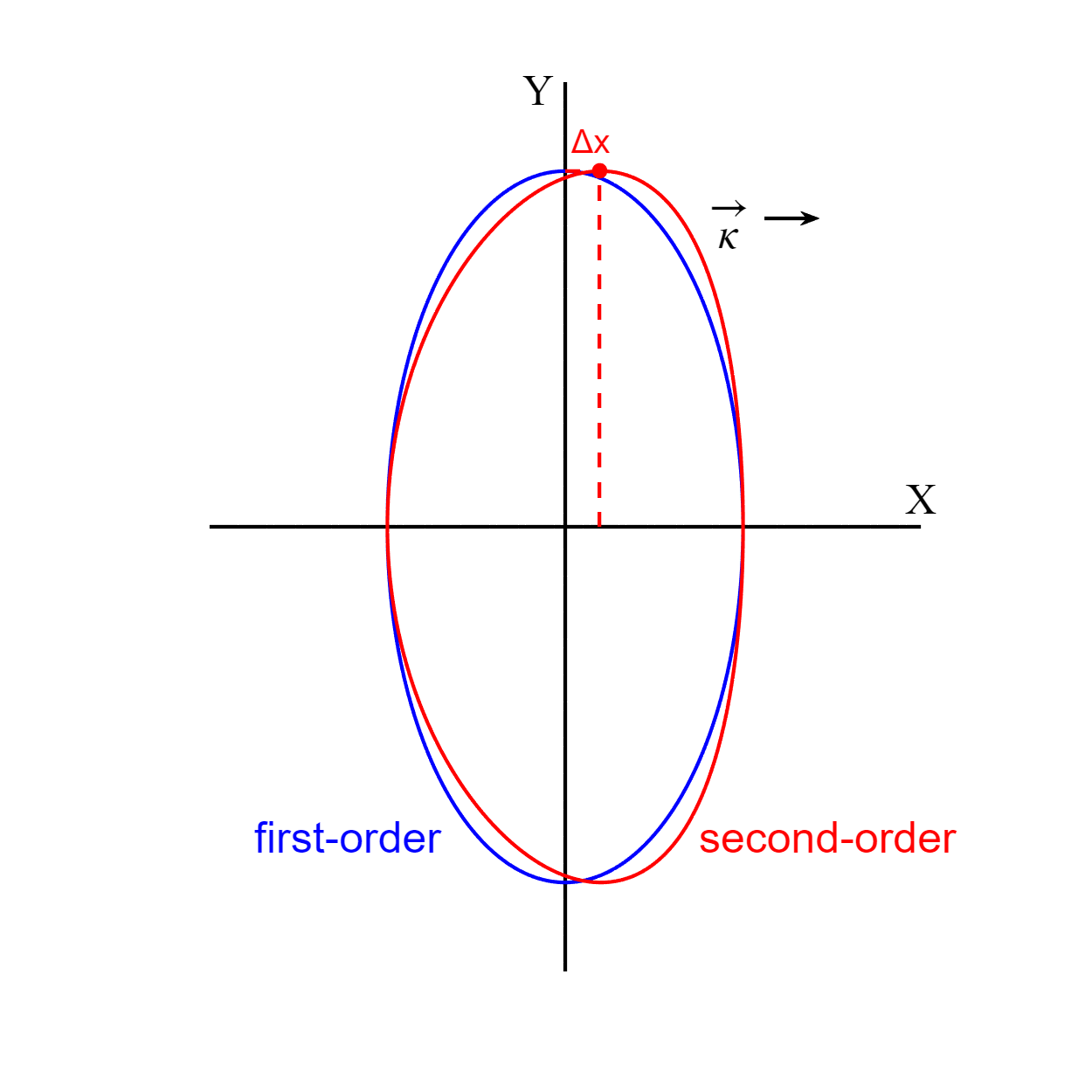} 
		\caption{}
		\label{ell.b}
	\end{subfigure}
    \caption{The cross sections of first- and second-order solutions.}
	\label{ell}
\end{figure}

\begin{figure}[!t]
	\centering
	\includegraphics[width=0.4\textwidth, keepaspectratio]{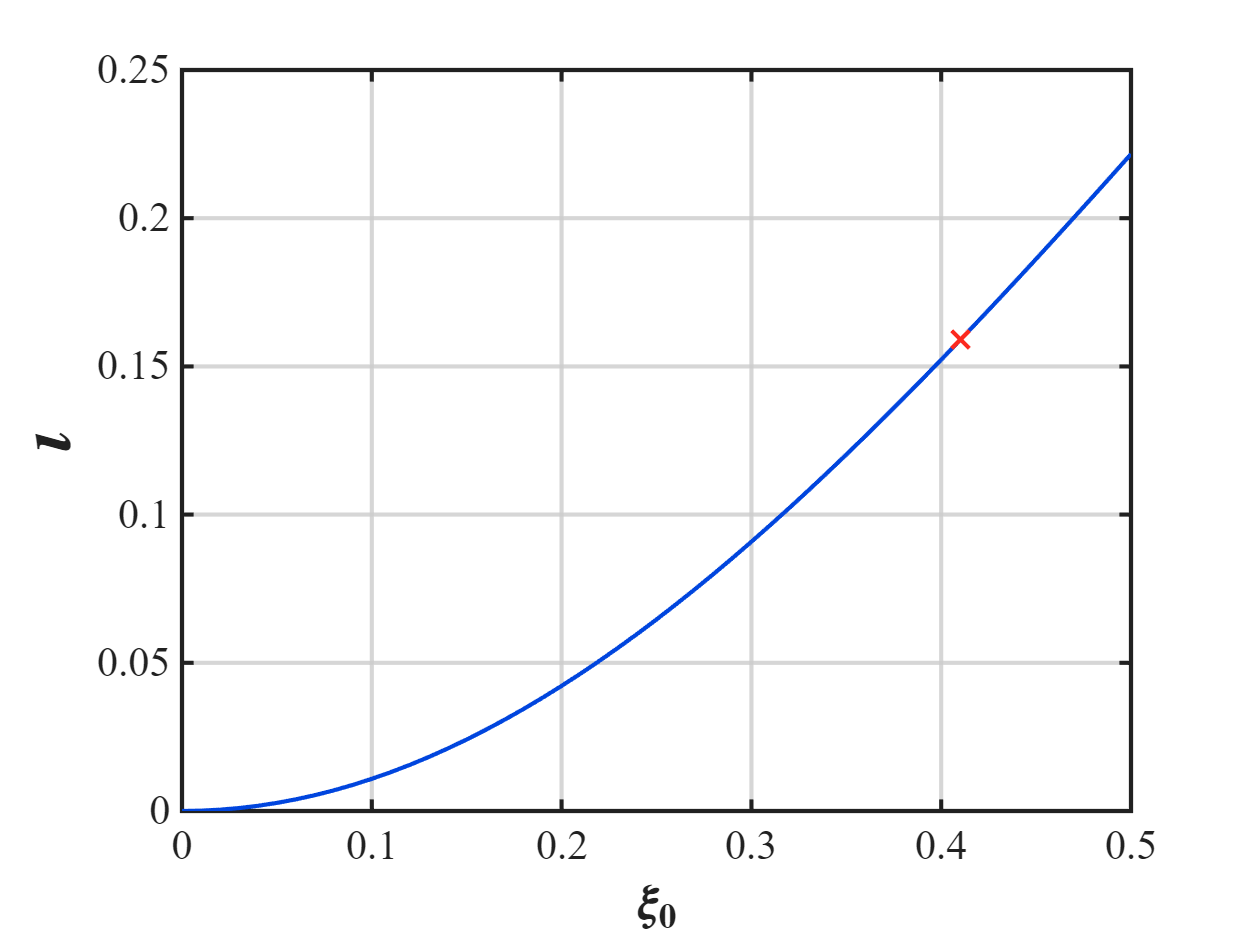} 	
	\caption{Rotational transform $\iota$ as a function of $\xi_0$.}
	\label{iota}
\end{figure}

\subsubsection{The second-order NAE solution and magnetic well}

When expanded to second order, the magnetic surface shape changes only slightly compared to the first-order result. Fig. 2b compares the first- and second-order results. The main second-order effect is a finite triangularity of the magnetic surface, i.e., a slight deformation of the elliptical shape, as shown in Fig. 2b. This triangular deformation induces a magnetic well as explained below. 

Within the second-order expansion, we have
\begin{equation}
\label{eq:second-order-field}
\begin{aligned}[b]
B(\psi,\theta,\varphi) &= B_0(\varphi) + r B_1(\theta,\varphi) + r^2 B_2(\theta,\varphi) \\
&\approx B_0(\varphi)(1 + \kappa X(\psi,\theta,\varphi)),
\end{aligned}
\end{equation}
and
\begin{equation}
\label{eq:magnetic-well-approx}
W = 2\frac{V}{\langle{B^2}\rangle}\frac{d}{dV}\langle{\frac{B^2}{2}}\rangle
\approx \frac{\langle{B^2}\rangle - \langle{B_0^2}\rangle}{\langle{B_0^2}\rangle}.
\end{equation}

We consider the contribution of a specific magnetic surface at $\varphi = \varphi_0$ to the magnetic well,
\begin{equation}
\label{eq:local-magnetic-well}
\begin{aligned}[b]
W_{\varphi_0} &\approx \frac{\langle{B^2(\varphi_0)}\rangle_{\theta} - \langle{B_0^2(\varphi_0)}\rangle_{\theta}}{\langle{B_0^2(\varphi_0)}\rangle_{\theta}} \\
&\approx \frac{B_0^2(\varphi_0)\langle{[1 + \kappa(\varphi_0) X(\psi,\theta,\varphi_0)]^2}\rangle_{\theta} - B_0^2(\varphi_0)}{B_0^2(\varphi_0)} \\
&= 2\kappa(\varphi_0) \langle{X(\psi,\theta,\varphi_0)}\rangle_{\theta} + \kappa^2(\varphi_0) \langle{X^2(\psi,\theta,\varphi_0)}\rangle_{\theta}.
\end{aligned}
\end{equation}

This indicates that for a cross-section with nonzero curvature of the magnetic axis, a positive contribution to the magnetic well requires the poloidal average of $X$ 
to be greater than zero. Specifically, the magnetic surface shape must exhibit a triangular deformation toward the curvature direction in the toroidal sections\cite{rodriguez2023magnetohydrodynamic}, 
similar to what is illustrated in Fig. 2b. For the up-down symmetric cross-sections in Fig. 4, the triangularity is defined as $\delta = \frac{\Delta x}{r_x}$,
where $\Delta x = X(\theta_{Y_{\max}})-X_{\mathrm{mid}}$, $X_{\mathrm{mid}}=(X_{\max}+X_{\min})/2$ is the midpoint of the $X$ extent, and $r_x=(X_{\max}-X_{\min})/2$. 
We prescribe the second-order parameter associated with the triangular deformation to be analogous to the profile of $\xi(\varphi)$, 
ensuring that the magnetic surfaces at the mirror sections remain circular. 
Therefore, our design strategy aims to induce a triangular deformation of the elliptical magnetic surfaces toward the curvature direction at the toroidal sections, thereby overcoming the magnetic hill generated by the mirror sections. This ensures the formation of an overall magnetic well. 

\begin{figure}[!t]
	\centering
	\begin{subfigure}[b]{0.4\textwidth}
		\includegraphics[width=\textwidth]{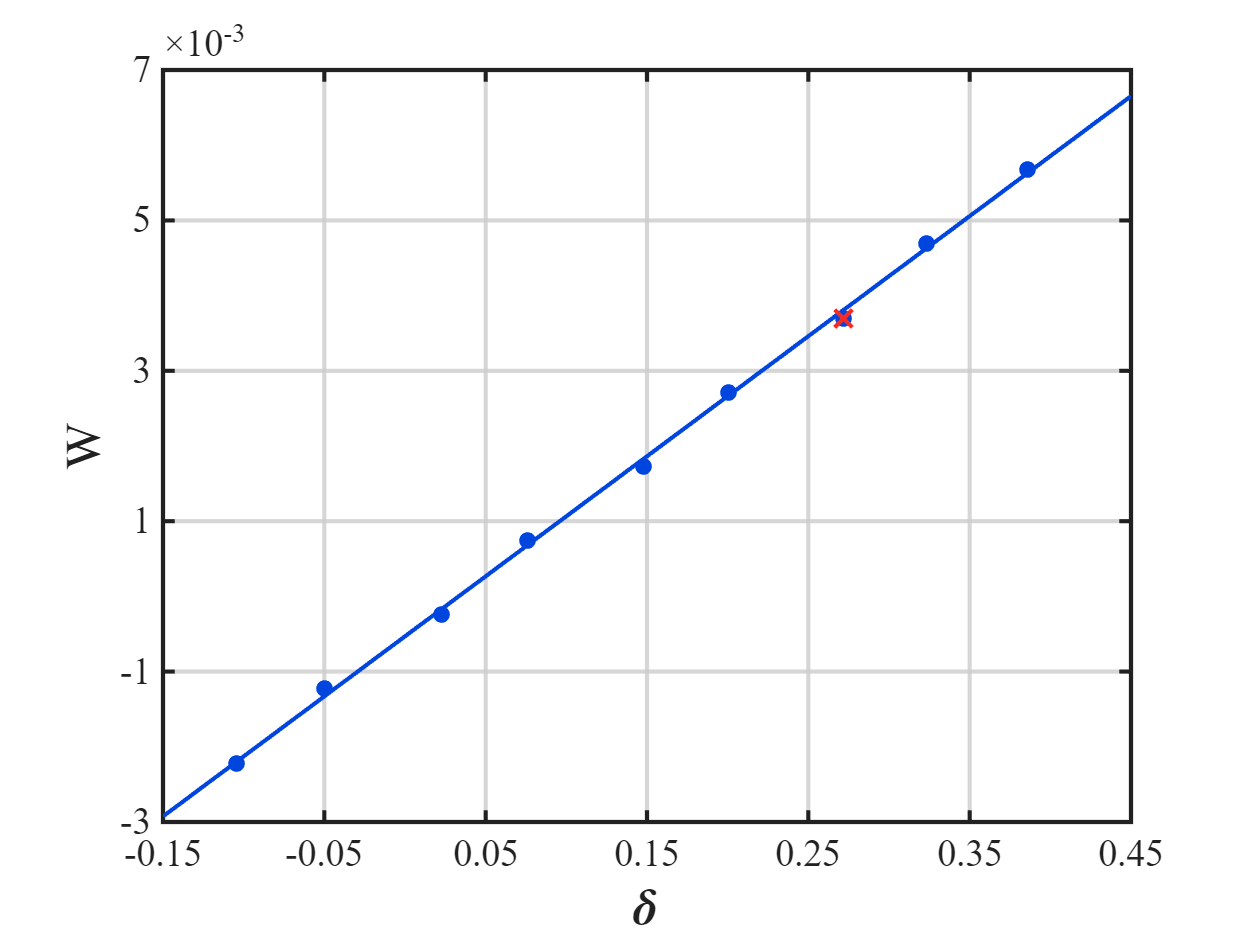} 
		\caption{}
		\label{tri.a}
	\end{subfigure}
	\quad
	\begin{subfigure}[b]{0.4\textwidth}
		\includegraphics[width=\textwidth]{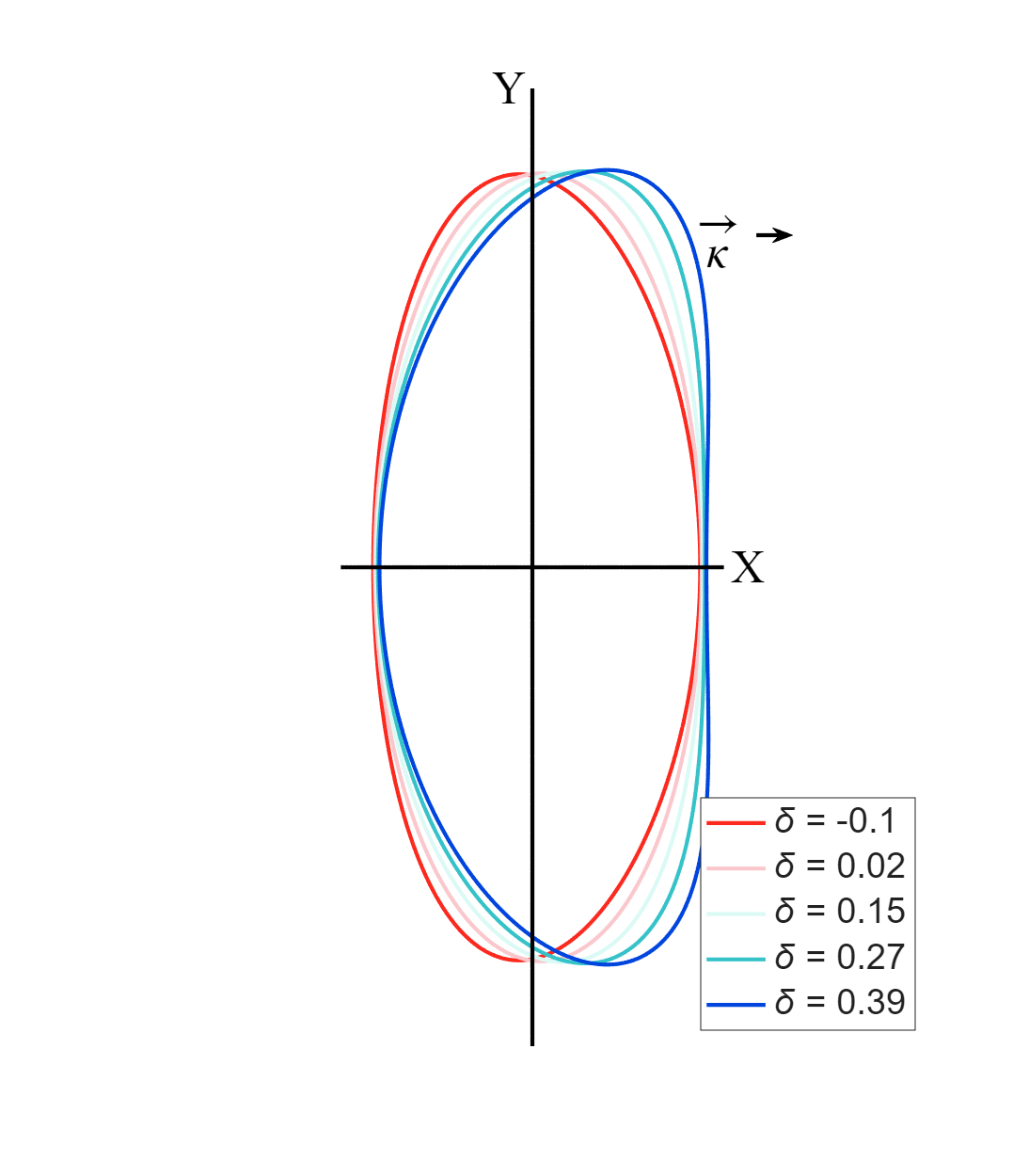} 
		\caption{}
		\label{tri.b}
	\end{subfigure}
    \caption{(a) Magnetic well $W$ versus $\delta$; (b) cross-sections for selected $\delta$ values.}
	\label{tri}
\end{figure}

To maintain the accuracy of the NAE, we perform the expansion up to a minor radius of $0.2\,\mathrm{m}$ at the mirror position. 
Fig. 4a plots the magnetic well $W$ as a function of triangularity $\delta$ of the magnetic surface at $\varphi = \frac{\pi}{2}$, 
while Fig. 4b presents the magnetic surface shapes for different values of $\delta$. 
We choose $\delta = 0.3$ as our design point, and the corresponding magnetic well is approximately $0.4\%$. 

In summary, we have designed an NAE equilibrium with a rotational transform of $\iota=0.16$
and a positive magnetic well. The shape of the last magnetic surface is shown in Fig. 5. 
The magnetic surface shapes at different toroidal angles are shown in Fig. 6. 

\begin{figure}[!t]
	\centering
	\includegraphics[width=0.5\textwidth, keepaspectratio]{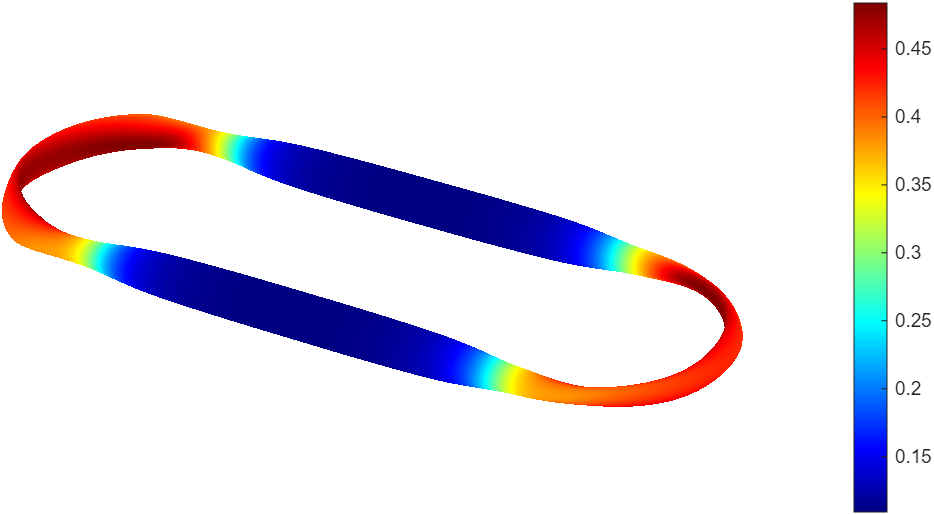} 	
	\caption{magnetic surface and magnetic field  of the LMD configuration.}
	\label{flux}
\end{figure}

\begin{figure}[!t]
	\centering
	\includegraphics[width=0.5\textwidth, keepaspectratio]{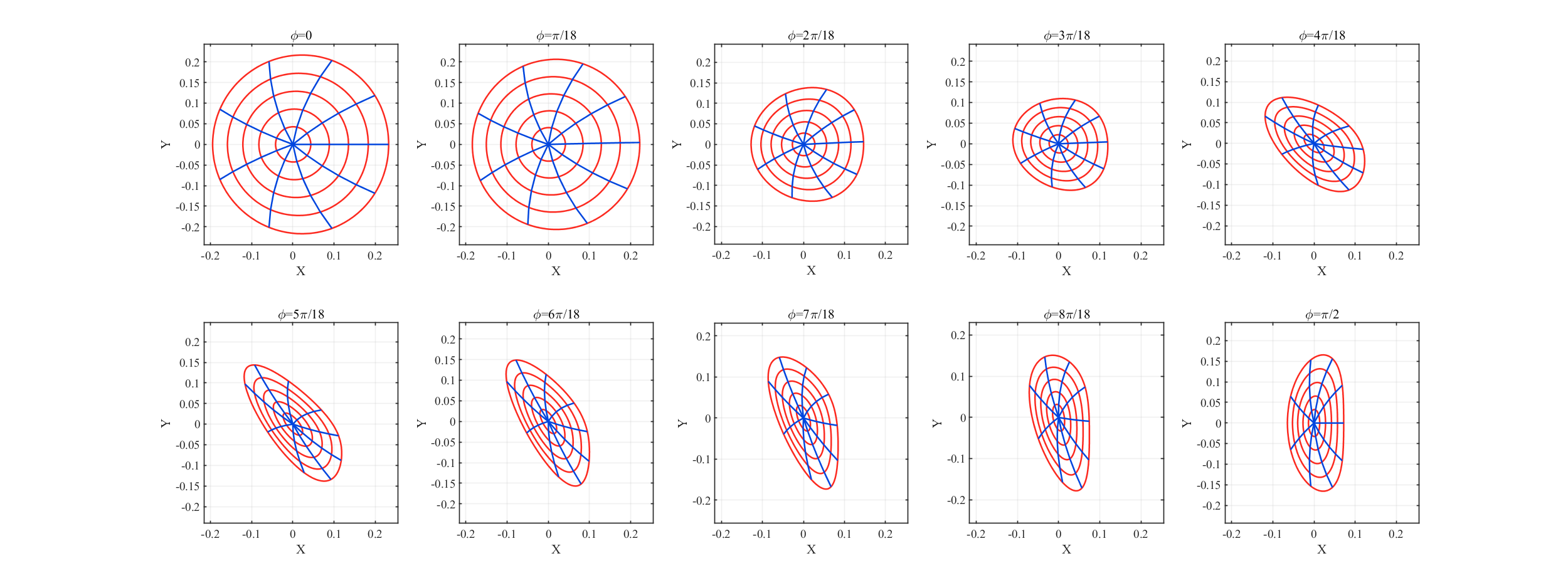} 	
	\caption{magnetic surface shape for different $\varphi$ values.}
	\label{flux_2d}
\end{figure}

\subsection{Equilibrium solutions with GVEC code}
Here we present results of equilibrium calculations using the GVEC code\cite{hindenlang2025computing}. The 3D fixed-boundary numerical equilibrium solutions are obtained using the NAE solution for the boundary magnetic surface as input. 
First, we obtain vacuum magnetic field from GVEC and use it to verify the NAE solution. Second, finite-beta equilibria are obtained using GVEC to investigate the effects of finite beta on the equilibrium.

\subsubsection{Vacuum solutions}

Fig. 7 compares the iota profile of the NAE equilibrium with that of the corresponding numerical equilibrium computed by GVEC. 
Compared with the NAE result, the iota profile calculated by GVEC is slightly nonuniform, but the two profiles are in good agreement. The small discrepancy is due to higher-order effects contained in the numerical equilibrium. 

\begin{figure}[!t]
	\centering
	\includegraphics[width=0.4\textwidth, keepaspectratio]{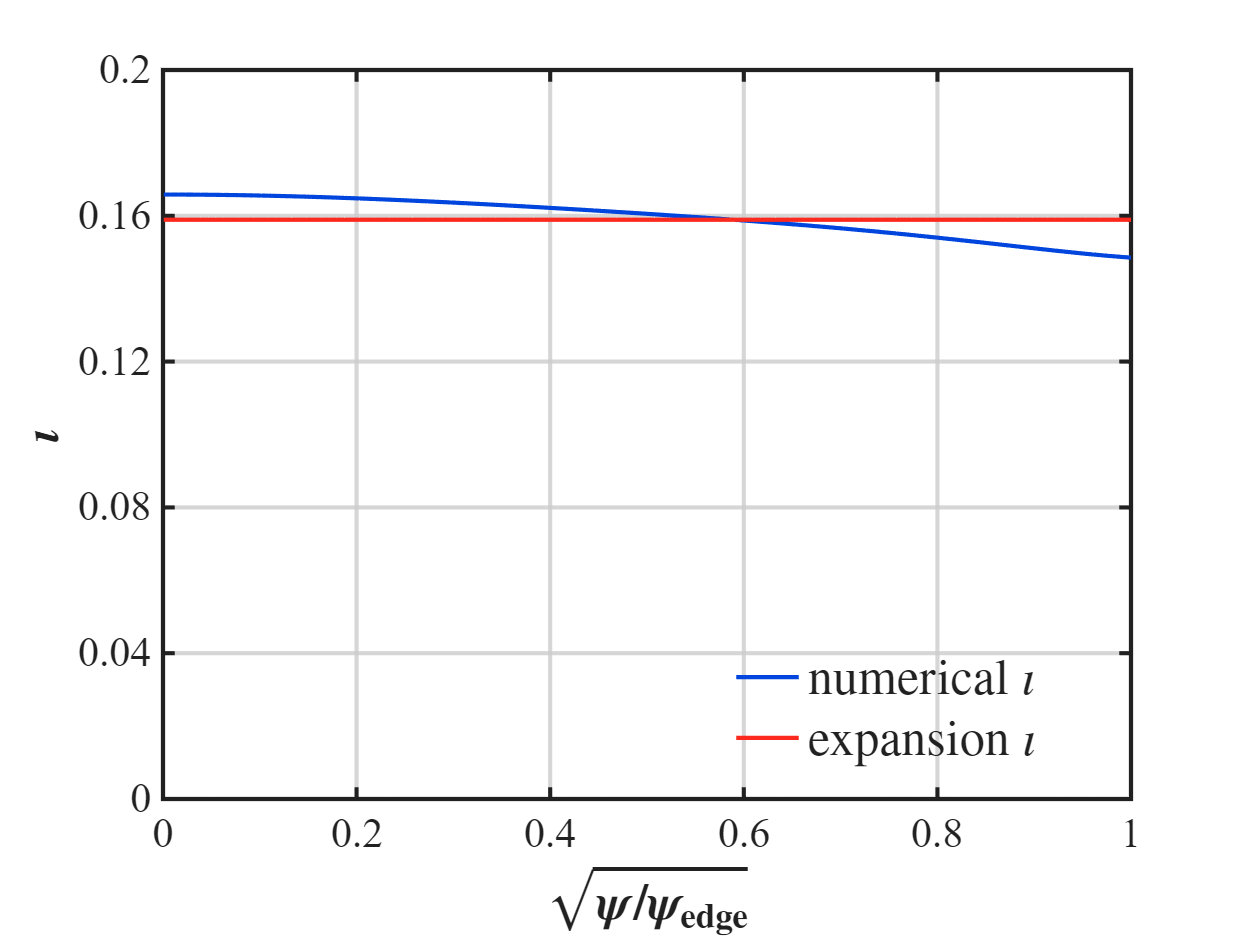} 	
	\caption{Comparison of rotational transform profiles between the GVEC calculation and the NAE solution.}
	\label{iota_pro}
\end{figure}

Since GVEC computes fixed-boundary equilibria, the last closed flux surface (LCFS) of the obtained equilibrium is the same as that of the NAE solution. Thus, we can also compare the magnetic axis location to assess the accuracy of the NAE equilibrium. 
Fig. 8 shows the discrepancy in magnetic axis position, normalized to the minor radius, at different toroidal angles. We observe that the discrepancy is very small, within $1\%$. 
This result indicates that the calculated location of magnetic axis is very close to that of NAE equilibrium. This comparison, together with the iota comparison, verifies the accuracy of the NAE equilibrium.  
\begin{figure}[!t]
	\centering
	\includegraphics[width=0.4\textwidth, keepaspectratio]{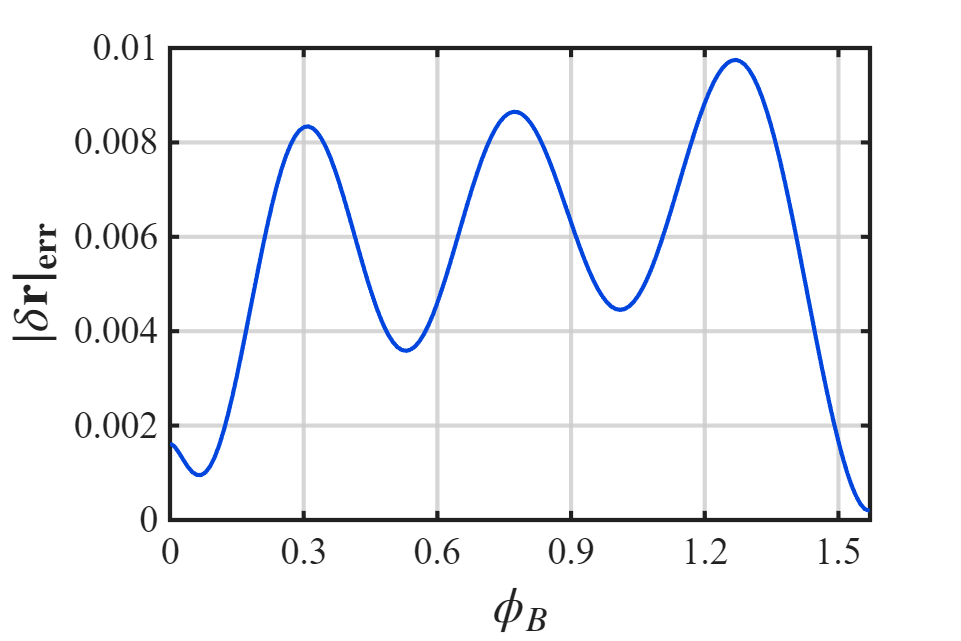} 	
	\caption{The relative position error normalized to the minor radius of LMD as a function of toroidal angle. }
	\label{r_err}
\end{figure}

Furthermore, the magnetic well calculated by GVEC is about $1\%$, 
which is larger than the NAE result. This confirms numerically that the magnetic well indeed exists for our NAE design of the linked mirror. It should be pointed out that the numerical magnetic well is about 2.5 times the NAE value. This sizable discrepancy can be understood from the fact that the well arises from second-order NAE effects and is presumably sensitive to small differences between the NAE and numerical equilibria.

\subsubsection{Finite beta equilibrium}

We now consider finite-beta equilibria with the pressure profile given by $P = p_0 (1 - s^2)^3$, 
where $s$ is the normalized toroidal magnetic flux. 
For simplicity, we define beta used throughout this paper as $\beta = \frac{p_0}{B_0^2(\varphi =0)/2\mu_0}$, 
i.e., the beta at the midpoints of the mirror sections.

We find that as beta increases, the magnetic axis gradually shifts outward as shown in Fig. 9. 
By analogy with tokamaks, we define the critical equilibrium beta as the beta value at which the magnetic axis shift equals half the minor radius. We have evaluated the critical beta for two different pressure profiles: 
it is approximately $15\%$ for the steeper profile of $P = p_0 (1 - s^2)^3$ and $25\%$ for the flatter profile of $P = p_0 (1 - s^2)$. 
The flatter pressure profile therefore yields a significantly higher critical beta. 
Fig. 10 shows the cross section of magnetic surfaces at the respective critical beta values for the steeper profile.

Furthermore, it is found that finite beta has a strong effect on the depth of the magnetic well.  
Fig. 11 shows that the magnetic well at the LCFS increases significantly as beta increases. This suggests that the stability of the interchange mode is strongly enhanced by finite-beta effects. 

\begin{figure}[!t]
	\centering
	\includegraphics[width=0.4\textwidth, keepaspectratio]{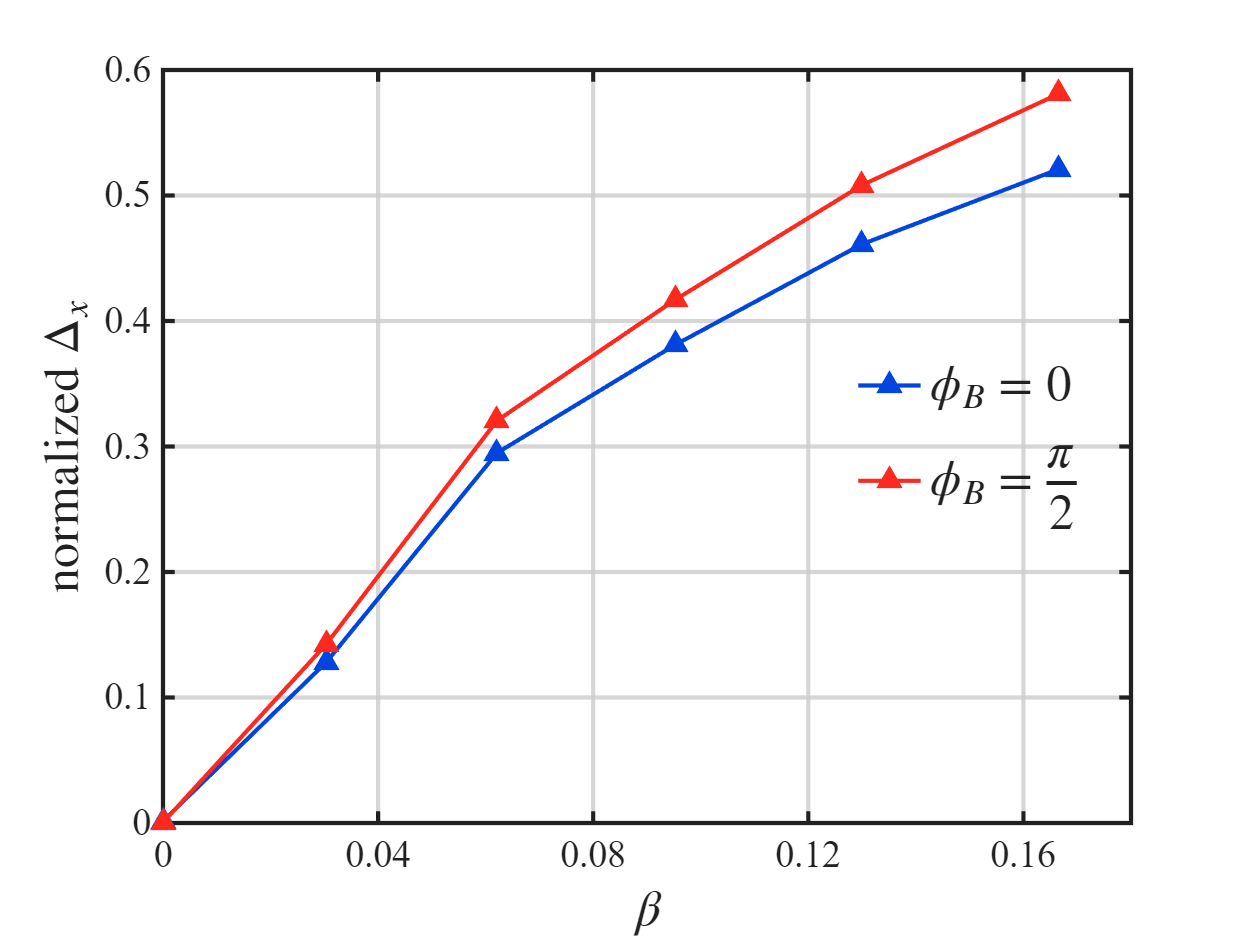} 	
	\caption{The Shafranov shift\cite{shafranov1963equilibrium} as a function of plasma beta. }
	\label{dx_beta}
\end{figure}

\begin{figure}[!t]
	\centering
	\begin{subfigure}[b]{0.4\textwidth}
		\includegraphics[width=\textwidth]{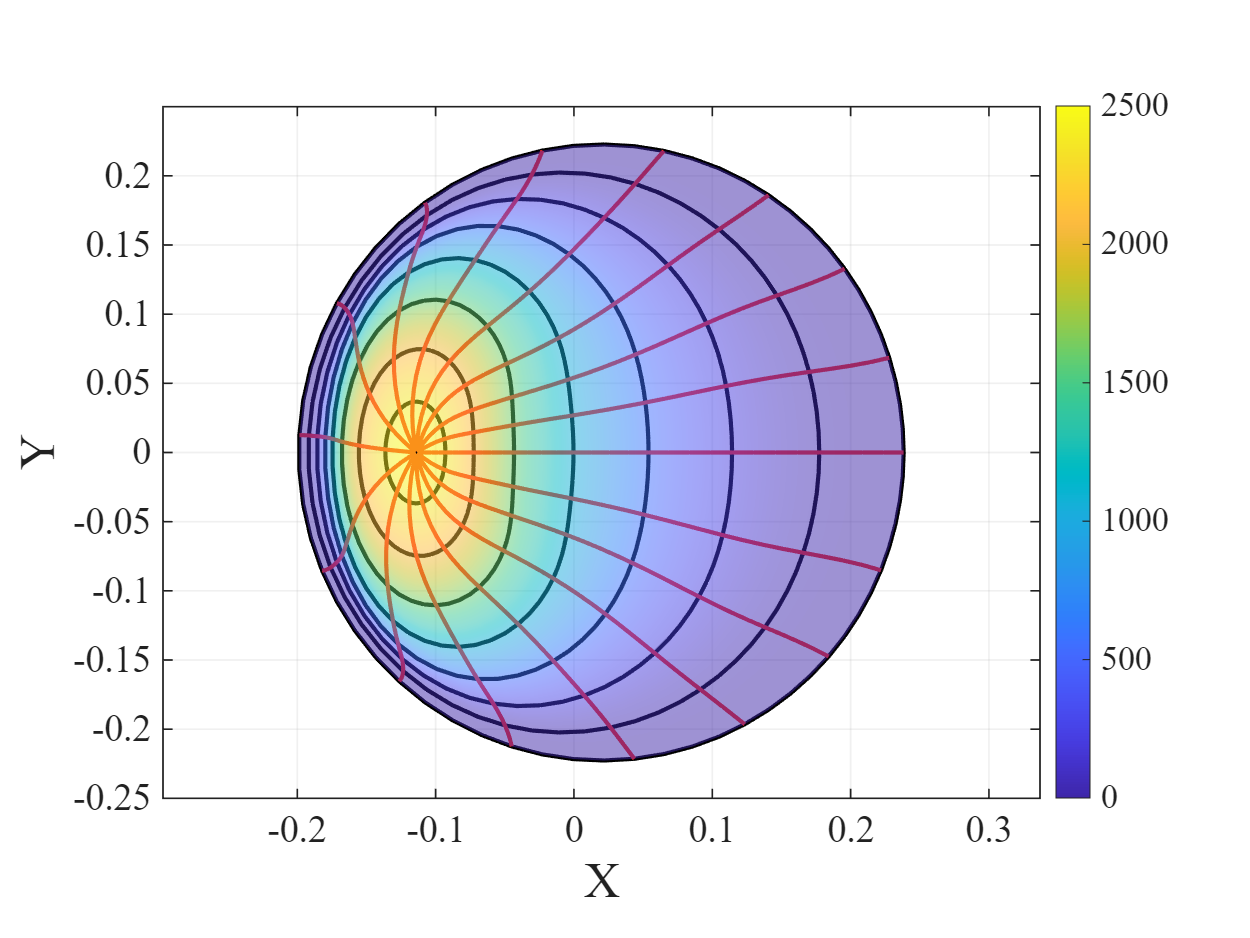} 
		\caption{}
		\label{cri.a}
	\end{subfigure}
	\quad
	\begin{subfigure}[b]{0.4\textwidth}
		\includegraphics[width=\textwidth]{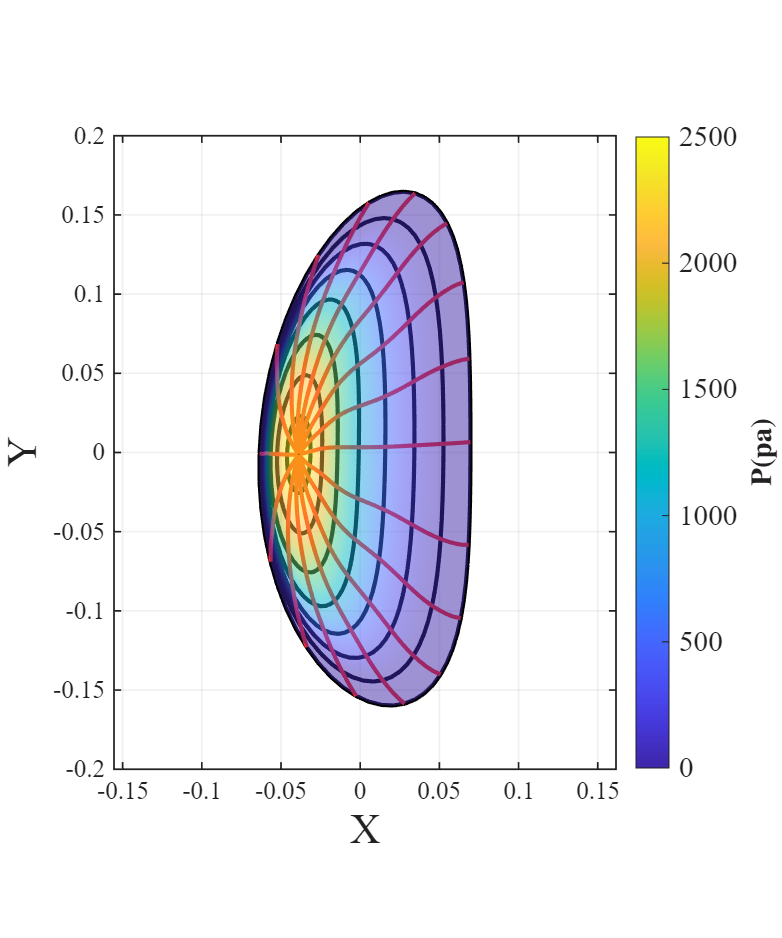} 
		\caption{}
		\label{cri.b}
	\end{subfigure}
    \caption{The cross section at (a) $\varphi=0$ and (b) $\varphi=\frac{\pi}{2}$ }
	\label{cri}
\end{figure}

\begin{figure}[!t]
	\centering
	\includegraphics[width=0.4\textwidth, keepaspectratio]{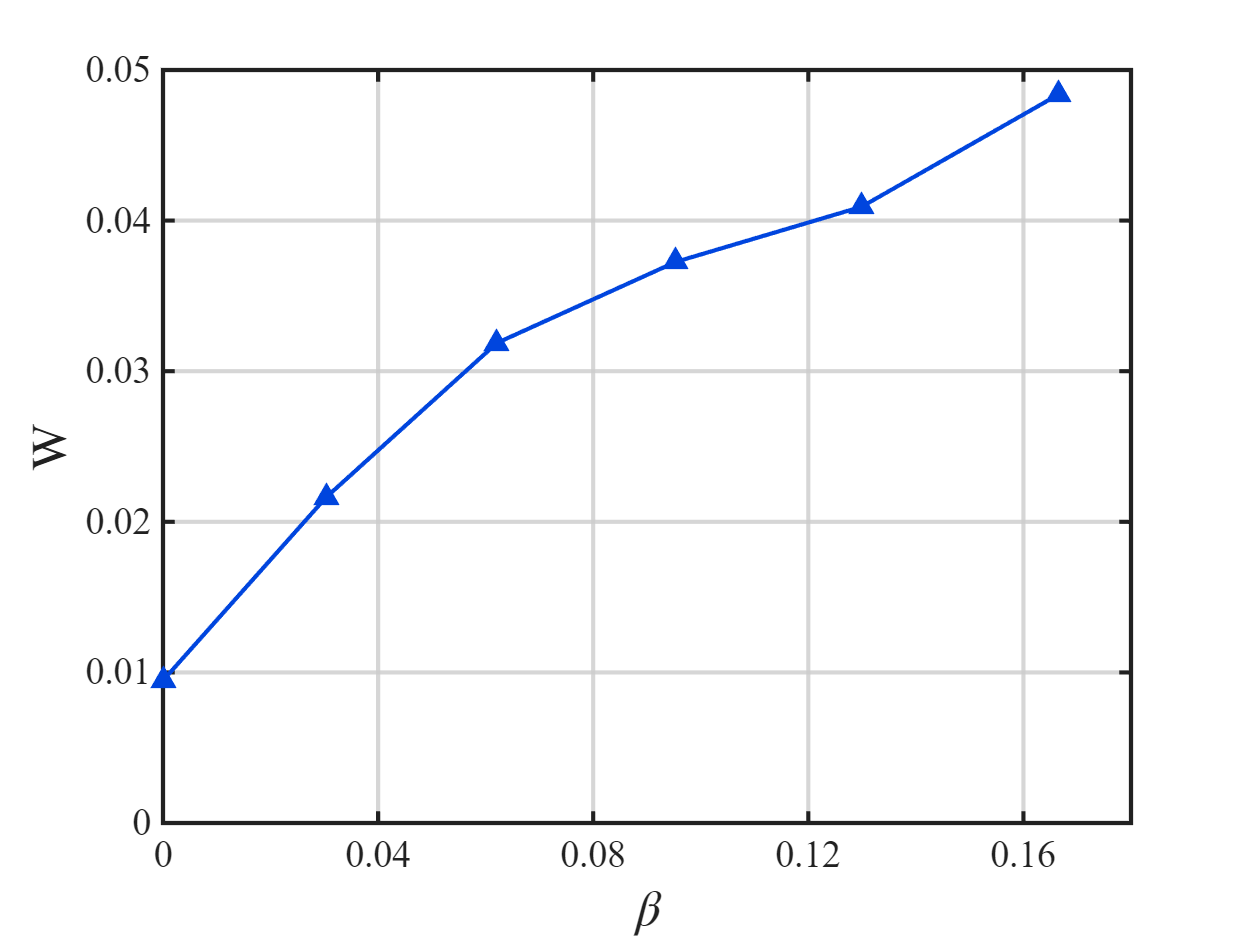} 	
	\caption{The magnetic well of the last closed flux surface as a function of beta. }
	\label{well_beta}
\end{figure}

\subsection{Coil design}
We use the FOCUS code\cite{zhu2018new} to design a coil set to reproduce the NAE vacuum magnetic field solution. The coil complexity is optimized through a loss function incorporating coil length, curvature, and torsion. 

The coil set consists of 56 coils: those at the mirror sections remain almost circular, while those at the toroidal sections have three-dimensional shapes. 
Fig. 12 shows the coils (in gold) and the LCFS (in purple) obtained by tracing magnetic field lines. 
Fig. 13 shows the distribution of normalized normal-field error at the target magnetic surface of the NAE equilibrium. The field error is mainly concentrated at the toroidal sections and is less than $1.3\%$. 
We compare the magnetic surface shapes obtained by tracing magnetic field lines from the coils against the near-axis equilibrium. 
Fig. 14 shows the comparison of the magnetic surfaces at $\varphi=0$ and $\varphi=\frac{\pi}{2}$ between the NAE target surface (red curve) and the surfaces (blue lines) generated by the coil set. We observe that the target surfaces closely match those from the coil set. It should be noted that the direction of the x-axis here is opposite to that of Fig. 10. 
\begin{figure}[!t]
	\centering
	\begin{subfigure}[b]{0.5\textwidth}
		\includegraphics[width=\textwidth]{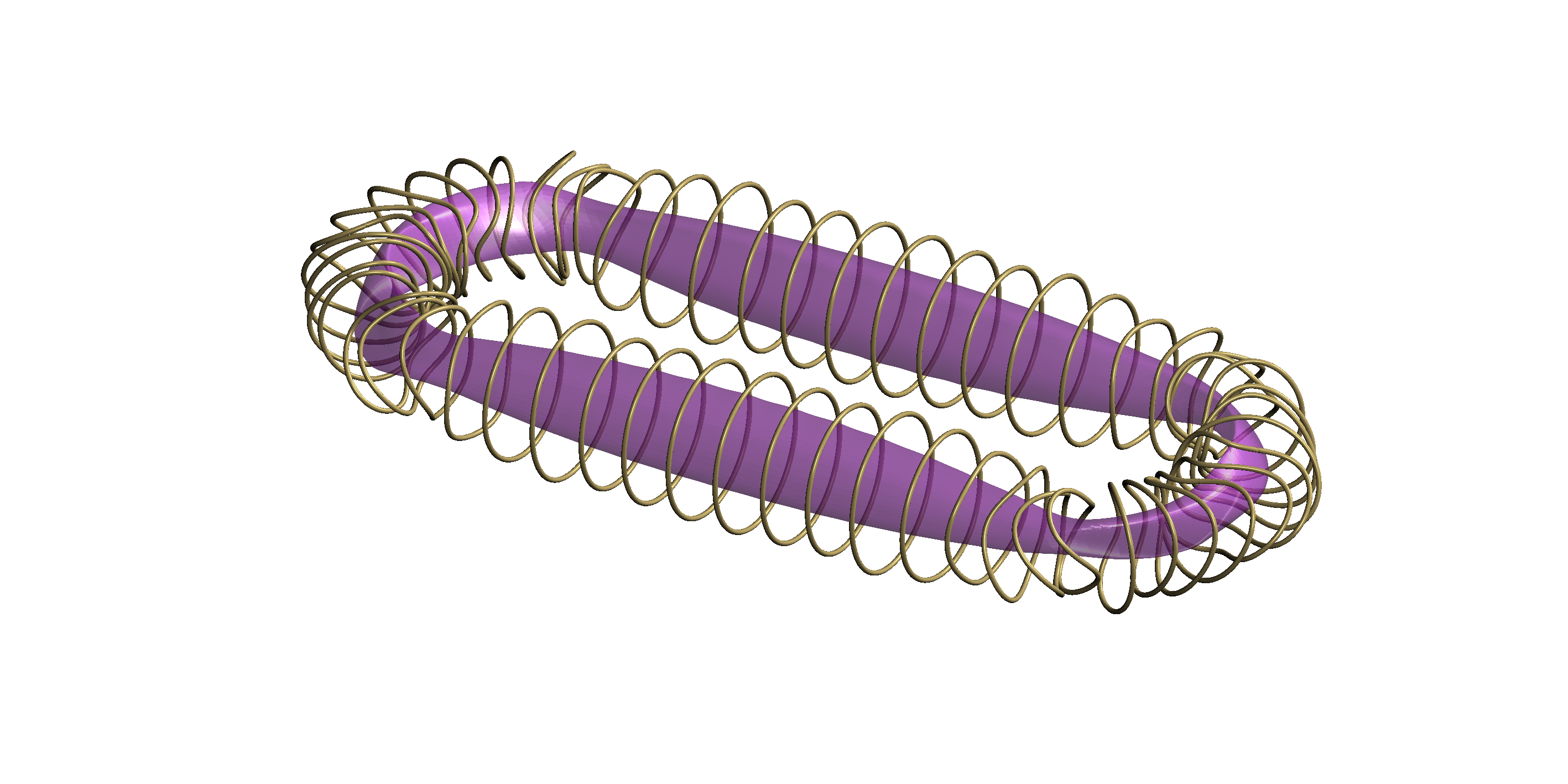} 
		\caption{3D view}
		\label{coil_surf.a}
	\end{subfigure}
	\quad
	\begin{subfigure}[b]{0.5\textwidth}
		\includegraphics[width=\textwidth]{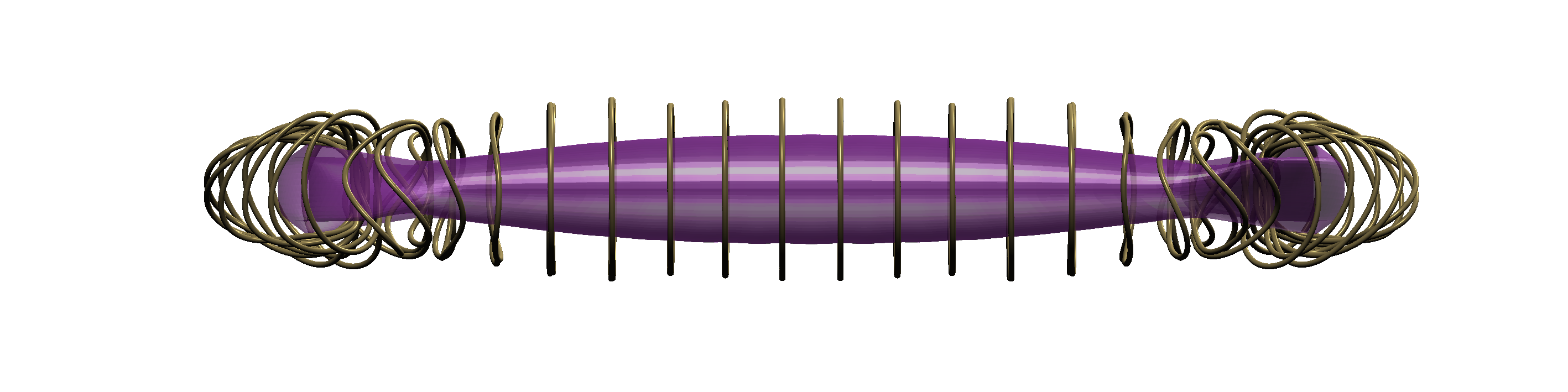} 
		\caption{Front view}
		\label{coil_surf.b}
	\end{subfigure} 	
	\quad
	\begin{subfigure}[b]{0.5\textwidth}
		\includegraphics[width=\textwidth]{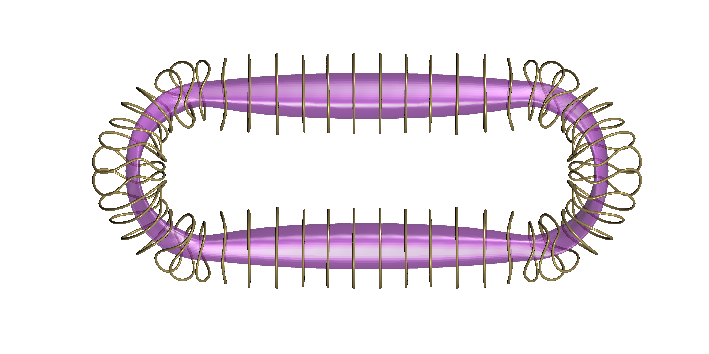} 
		\caption{Top view}
		\label{coil_surf.c}
	\end{subfigure}
	\caption{The coil configuration (in gold) and the LCFS (in purple). }
	\label{coil_surf}
\end{figure}

\begin{figure}[!t]
	\centering
	\includegraphics[width=0.4\textwidth, keepaspectratio]{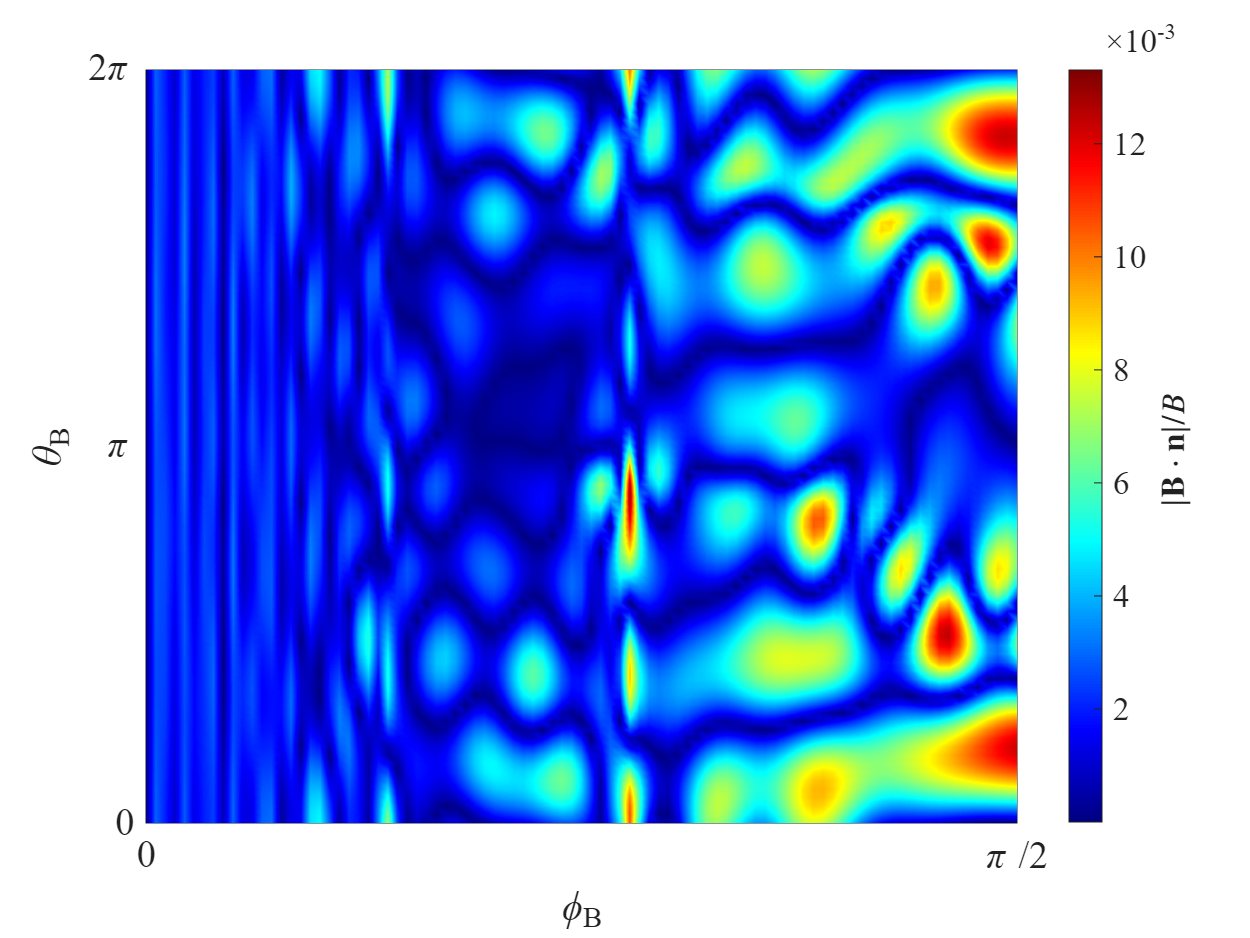} 	
	\caption{Normal-magnetic-field error distribution on the target magnetic surface.}
	\label{Bn}
\end{figure}

\begin{figure}[!t]
	\centering
	\begin{subfigure}[b]{0.4\textwidth}
		\includegraphics[width=\textwidth]{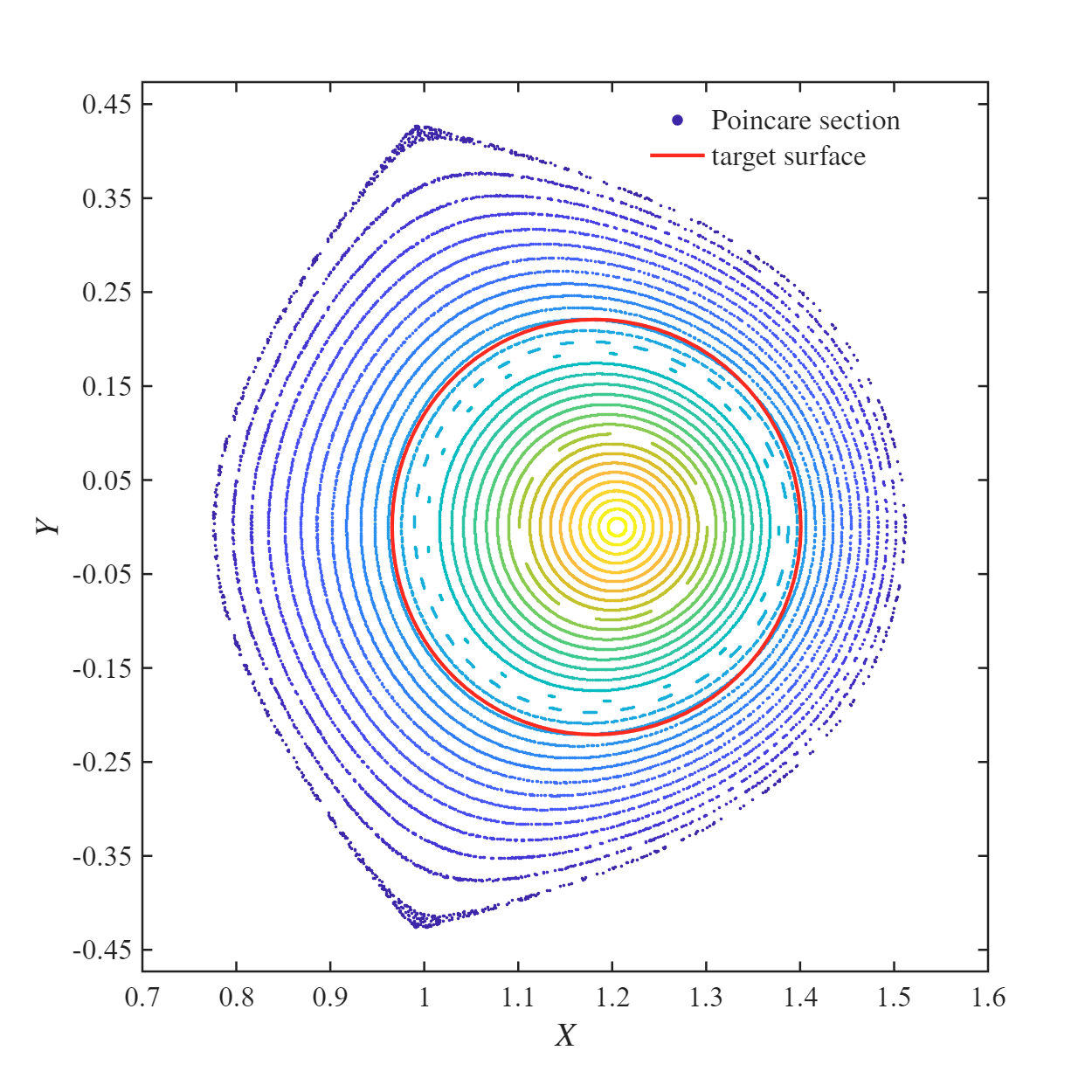} 
		\caption{}
		\label{coil_flux.a}
	\end{subfigure}
	\quad
	\begin{subfigure}[b]{0.4\textwidth}
		\includegraphics[width=\textwidth]{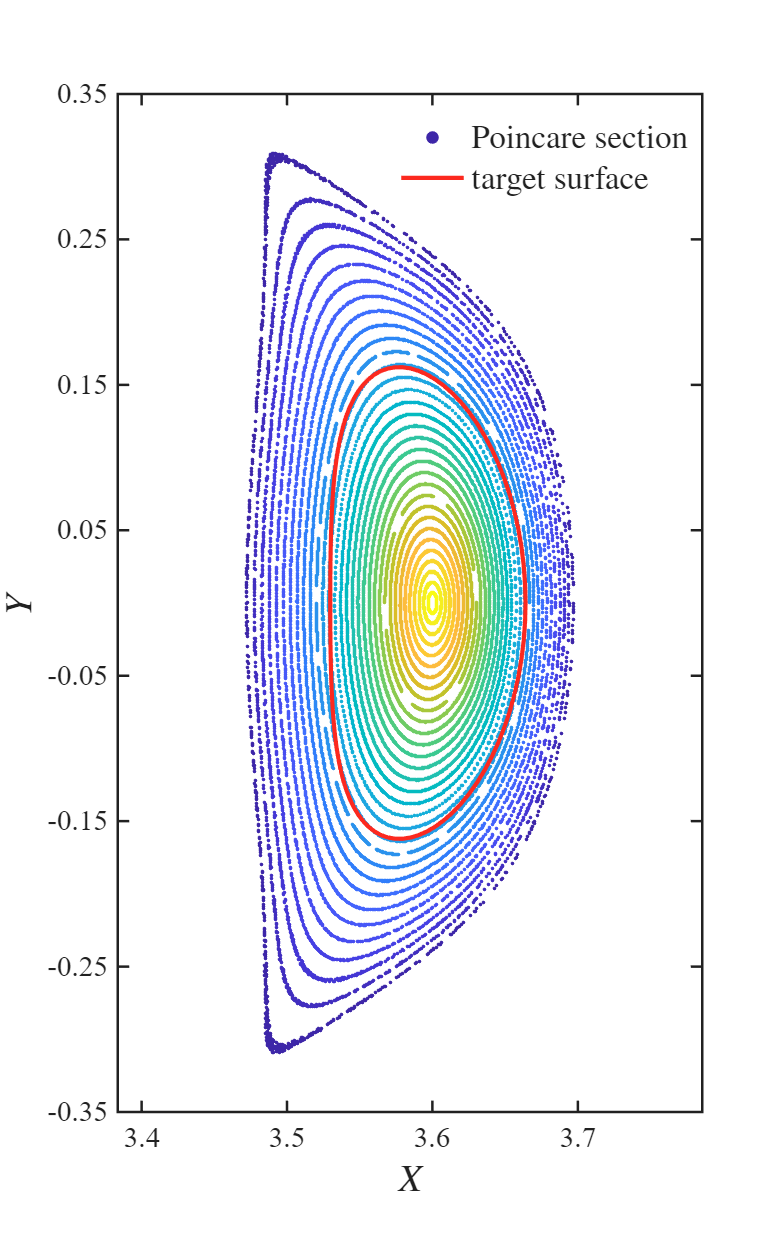} 
		\caption{}
		\label{coil_flux.b}
	\end{subfigure}
    \caption{the comparison of the magnetic surfaces obtained from coil configuration and near-axis expansion at (a) $\varphi=0$ and (b) $\varphi=\frac{\pi}{2}$ }
	\label{coil_flux}
\end{figure}

The rotational transform profile is also calculated from  the designed coil set and agrees well with the NAE equilibrium result. Furthermore, the magnetic well calculated from the coil set is $0.2\%$, which is also comparable to the magnetic well of the NAE equilibrium. 
In summary, a coil set has been designed that reproduces the NAE vacuum magnetic field and preserves the magnetic well.

\subsection{Particle confinement}
Using the above vacuum magnetic field from the coil set, we have carried out test-particle simulations to evaluate the collisionless loss. 
We use the test-particle code FP3D for this purpose. The main parameters include minor radius at the mirror sections $a_0 \approx 0.2\,\mathrm{m}$, magnetic field $B=0.1\,\mathrm{T}$, main ion energy $E = 100\,\mathrm{eV}$, which corresponds to a normalized ion gyroradius of $\rho/a_0 = 0.04$. A total of 10,000 test particles are used in the simulations.  Fig. 15 shows the collisionless particle confinement for mirror ratios of 4 and 8. The particle confinement fraction is $93\%$ for a mirror ratio of 8, indicating favorable confinement under the simulated conditions. It should be pointed out that a mirror ratio of 8 is included here in addition to the case of mirror ratio of 4 in order to investigate the effect of mirror ratio on particle confinement.
The particle losses are mainly from barely trapped particles, as shown in Fig. 16. These results are similar to those of the original linked-mirror design\cite{feng2021proposal}. 

\begin{figure}[!t]
	\centering
	\includegraphics[width=0.4\textwidth, keepaspectratio]{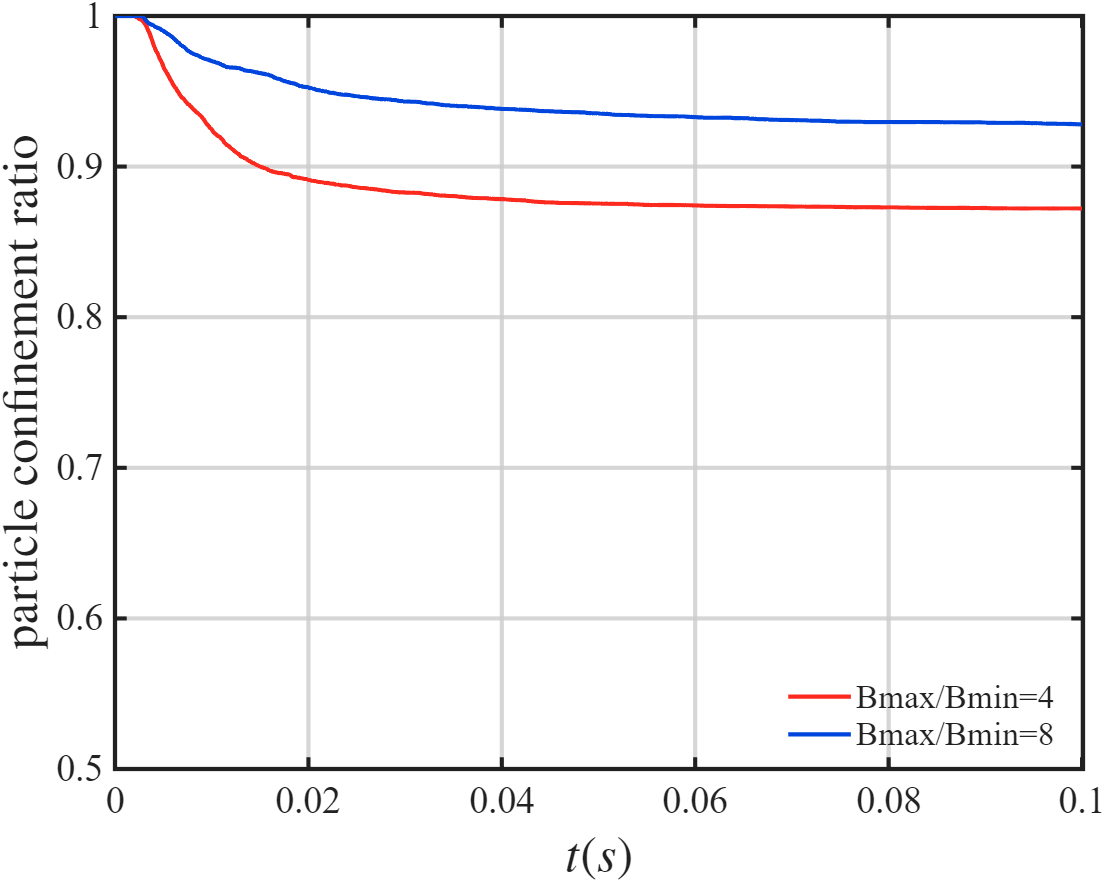} 	
	\caption{The collisionless particle confinement ratio of two coil configurations with mirror ratios of 4 and 8. }
	\label{coil_particle}
\end{figure}

\begin{figure}[!t]
	\centering
	\includegraphics[width=0.4\textwidth, keepaspectratio]{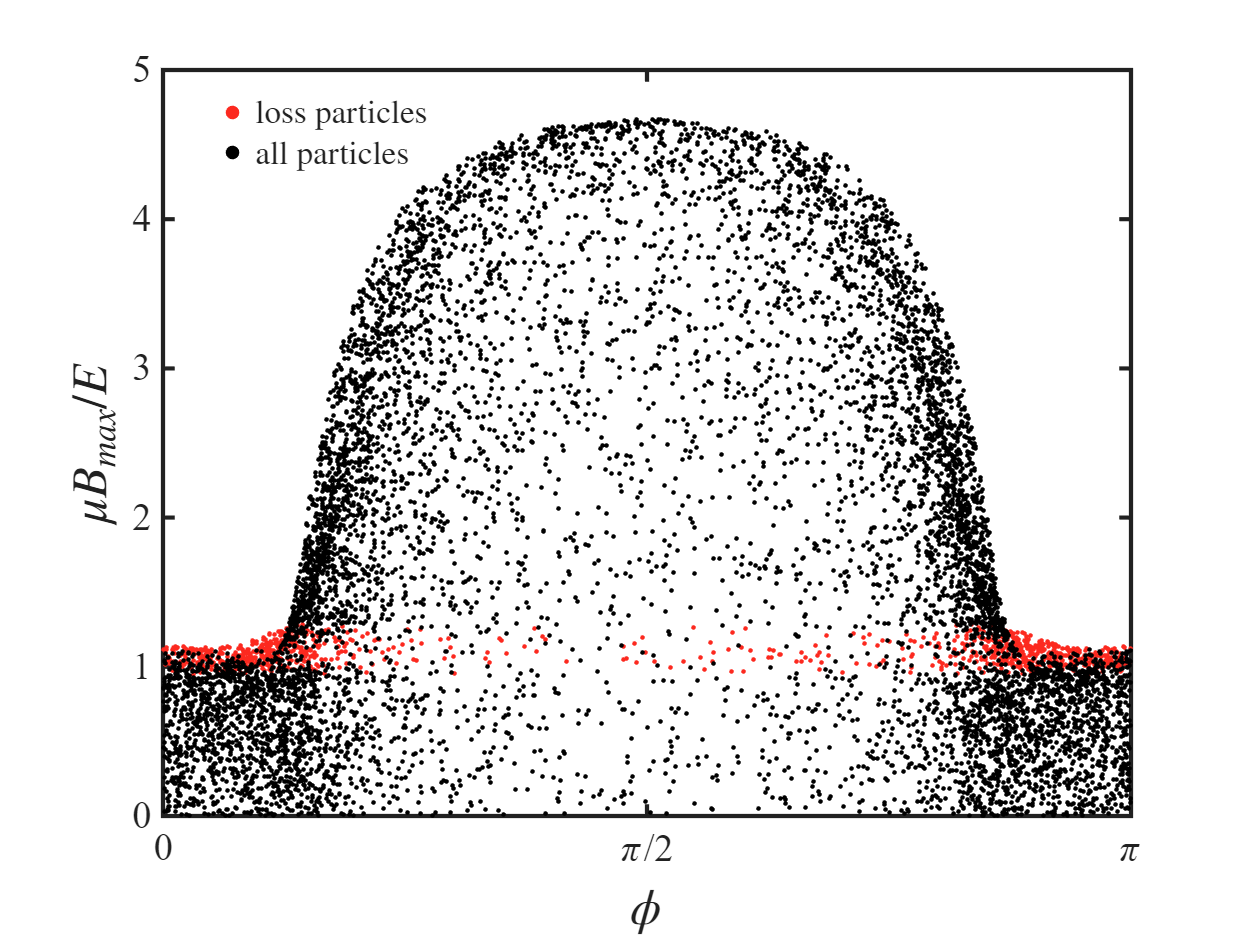} 	
	\caption{Distribution of confined particles and lost particles in $\mu B_{max}/E$ and $\phi$ space. }
	\label{particle_dis}
\end{figure}

\section{Discussions and conclusions}

We adopted the NAE method and used specially formulated expansion coefficients 
with respect to the toroidal angle to ensure that the magnetic structures of the two mirror sections remain straight and cylindrically symmetric. A finite rotational transform is generated by the rotation of elliptical magnetic surfaces at the toroidal connection sections. Furthermore, a magnetic well is formed due to the triangular shaping of the magnetic surfaces. 
Finally, a three-dimensional NAE equilibrium with a magnetic well structure is obtained, which guarantees stability against interchange modes.  

The three-dimensional magnetohydrodynamic (MHD) equilibrium code GVEC is used to verify the results of the NAE solutions for the vacuum magnetic field. The results show that the calculated rotational transform and magnetic axis position are consistent with the NAE results, verifying the accuracy of the NAE equilibria. 
Furthermore, the GVEC equilibrium also possesses a positive magnetic well with its value comparable to that of the NAE result. In addition, we have obtained a numerical equilibrium of a 3D linked mirror at finite beta. 
The results indicate that the magnetic well increases with increasing beta, which is beneficial for MHD stability. 
Finally, we have performed a preliminary estimation of the equilibrium critical beta, and the results show that the critical beta is substantial and is on the order of $20\%$. 

Finally, we have used the FOCUS code to design a coil set which reproduces the optimized NAE solution with respect to magnetic surfaces, rotational transform and magnetic well. This result shows that a relatively simple coil set can be found to reproduce and verify the NAE solution.

In conclusion, we have designed and verified a racetrack-shaped linked-mirror device for the magnetic confinement of plasma. It solves the long-standing problems of particle end loss and MHD instability in a traditional magnetic mirror, while its coil set is substantially simpler than that of a stellarator. The new design provides a novel solution for fusion reactors. Future work will consider further optimization of MHD stability and particle confinement, including neoclassical transport effects and further simplification of the 3D coils.

\section{Methods}
Here, we introduce the methods and codes employed in our design process. 
The first subsection describes the NAE method of equilibrium design in Boozer coordinates, while the second subsection presents the codes used. 

\subsection{Near-axis expansion method}

In Sec. II A, we have briefly described the NAE method employed in our design of linked mirror. Here, we describe the method in more detail. 

\subsubsection{Near-axis expansion in Boozer coordinate}

Starting from a prescribed magnetic axis, we expand the position vector of magnetic surfaces from the magnetic axis order by order (see Eq. 1). 
The magnetic field $B$  is expanded in the same way, but with an $r^0$ term included:
\begin{align}
    B(\psi,\theta,\varphi)&=B_0(\varphi) + r B_1(\theta,\varphi) + r^2 B_2(\theta,\varphi)+\cdots
\end{align}
The radial profiles $G(r)$, $I(r)$, $\iota(r)$ and $p(r)$ can be expanded in $r^2$. For example, 
\begin{align}
    p(\psi)&=p_0+r^2 p_2+r^4 p_4+\cdots
    \label{eq:pressure-expansion}
\end{align}
for the pressure profile. 

In Boozer coordinates, which are straight-field-line coordinates, the contravariant form and the covariant form of the magnetic 
field can be written:\cite{garren1991magnetic,landreman2018direct,landreman2019direct,plunk2019direct} 
\begin{align}
  &\textbf{B}_{con} = \nabla\psi \times \nabla\theta + \iota(\psi) \nabla\varphi \times \nabla\psi \\
    &\textbf{B}_{cov} = G(\psi)\nabla\varphi + I(\psi)\nabla\theta + \beta(\psi,\theta,\varphi) \nabla\psi 
\end{align}
The two representations of the magnetic field imply the following constraint: 
\begin{align}
 \textbf{B}_{con} = \textbf{B}_{cov}  
 \label{eq:field-constraint}
\end{align}

The plasma equilibrium equation is given by
\begin{align}
    (\nabla \times \textbf{B}_{cov}) \times \textbf{B}_{con} &= \mu_0 \nabla p 
    \label{eq:force-balance}
\end{align}

The radial expansions of position vector (Eq. 1), the magnetic field and the radial profiles of $G(r)$, $I(r)$, $\iota(r)$ and $p(r)$ are substituted into Eqs.~\eqref{eq:field-constraint} and~\eqref{eq:force-balance}, which are solved order by order. Details of the NAE method can be found in Ref.~\cite{garren1991magnetic}. For simplicity, our NAE design is performed under the conditions of zero pressure and vacuum equilibrium. 

\subsubsection{Magnetic Well} 
A magnetic well is known to provide a stabilizing effect on interchange modes in toroidal equilibria and
can serve as a simplified measure of MHD stability\cite{rodriguez2023magnetohydrodynamic}. The magnetic well is defined as follows:
\begin{align}
   W=2\frac{V}{\langle{B^2}\rangle}\frac{d}{dV}\langle{\frac{B^2}{2}}\rangle
\end{align}
where $V$ denotes the plasma volume within a magnetic surface and the angle brackets represent the magnetic-surface average. 

In an NAE equilibrium, the magnetic well can be expressed as\cite{landreman2020magnetic}:
\begin{align}
   W &= 2r^2 \left( \int_0^{2\pi} \frac{d\varphi}{B_0^2} \right)^{-1} \int_0^{2\pi} 
    \frac{d\varphi}{B_0^4} \left[ -\frac{3}{4} \left( B_{1s}^2 + B_{1c}^2 \right) + 
   B_{20}B_0 \right] \nonumber \\
   &+ \frac{\mu_0 r^2 p_2}{\pi} \int_0^{2\pi} \frac{d\varphi}{B_0^2}
\end{align}
Here $B_{1s}(\varphi)$ and $B_{1c}(\varphi)$ are the coefficients of $\sin\theta$ and $\cos\theta$ in $B_1(\theta,\varphi)$, respectively,
and $B_{20}(\varphi)$ is the component of $B_2(\theta,\varphi)$ independent of $\theta$.
The coefficient $p_2$ is the second-order coefficient defined in Eq.~\eqref{eq:pressure-expansion}. $\mu_0$ is the vacuum permeability.

\subsection{The codes used} 
\subsubsection{GVEC code}

GVEC\cite{hindenlang2025computing} builds upon the ideas of the well-established 3D equilibrium code VMEC. However, VMEC can only solve equilibria in the cylindrical coordinates $(R,\varphi,Z)$. In contrast, 
GVEC allows calculations in curvilinear coordinates that are constructed from two linearly 
independent vectors attached to a guiding curve referred to as the "G-frame" (general coordinate frame). 
The G-frame allows us to represent the geometry of stellarators of essentially any shape with closed 3D surfaces. 
Both the cylindrical coordinate system and the Frenet-Serret frame can be used within the G-frame.  

In GVEC, the geometry is described by 2D cross-sections in two coordinates $(q^1,q^2)$, and parameterized by a third periodic
coordinate $\zeta$. In the cylindrical coordinates,  
$\textbf{r}(\theta,\phi)=R(\theta,\phi)\textbf{e}_R(\phi)+Z(\theta,\phi)\textbf{e}_Z(\phi)$, 
so that $(q^1=R,q^2=Z,\zeta=\phi)$. 
In the Frenet-Serret frame, $\textbf{r}(\theta,\zeta) = \textbf{r}_{0}(\zeta) + q^1(\theta,\zeta) \textbf{n}(\zeta) 
+ q^2(\theta,\zeta) \textbf{b}(\zeta)$. For the same guiding curve $\textbf{r}_{0}(\zeta)$, $\zeta$ can be defined arbitrarily:
it can be defined either as the curve length $s$ or the Boozer toroidal angle $\varphi$. 
Because our design includes the straight mirror segments, the linked-mirror equilibria cannot be solved using VMEC, since cylindrical coordinates are not appropriate for describing the geometry of straight mirror sections. On the other hand, the local coordinates based on the Frenet-Serret frame are suitable for calculating linked-mirror equilibria. Indeed, we have successfully used GVEC to calculate
linked-mirror equilibria. 

\subsubsection{FOCUS code}
FOCUS\cite{zhu2018new} is a numerical code for the design of stellarator coils without requiring a toroidal winding surface. 
It represents a 3D coil geometry using a Fourier series. A closed coil curve is  generally represented in Cartesian 
coordinates as $\textbf{x}(t)=x(t)\textbf{i} + y(t)\textbf{j} + z(t)\textbf{k}$ where x(t), y(t) and z(t) are given by 
\begin{align}
   x(t) = x_{c,0} + \sum_{n=1}^{N_f} \left[ x_{c,n} \cos(nt) + x_{s,n} \sin(nt) \right] 
\end{align}
\begin{align}
   y(t) = y_{c,0} + \sum_{n=1}^{N_f} \left[ y_{c,n} \cos(nt) + y_{s,n} \sin(nt) \right] 
\end{align}
\begin{align}
   z(t) = z_{c,0} + \sum_{n=1}^{N_f} \left[ z_{c,n} \cos(nt) + z_{s,n} \sin(nt) \right] 
\end{align}
with $t \in \left[ 0, 2\pi \right]$. The shape of a coil is then fully determined by the $3 \times (2N_f+1)$ Fourier coefficients. 

The coil parameters are varied to minimize a target function that incorporates both physics and engineering objective functions, 
\begin{align}
   \chi^2(\mathbf{X}) = \sum_j w_j \left( \frac{f_j(\mathbf{X}) - f_{j,o}}{f_{j,o}} \right)^2 
\end{align}
where $f_j(\mathbf{X})$ is the $j$th objective function, $f_{j,o}$ denotes the desired value of the $j$th target, and $w_j$ is the corresponding weight coefficient. 
The most important objective function is the normal-field error on the plasma boundary; it can be expressed as 
\begin{align}
f_B(\mathbf{X}) \equiv \int_{\mathcal{S}} \frac{1}{2} \left( \mathbf{B} \cdot \mathbf{n} \right)^2 dS
\end{align}
In vacuum, $B$ can be calculated directly using the Biot-Savart law,
\begin{align}
\mathbf{B} = \frac{\mu_0}{4\pi} \sum_{i} I_i \oint_{C_i} \frac{d\mathbf{l}_i \times \mathbf{r}}{r^3}.
\end{align}
Here, $i$ denotes the $i$th coil. Other objective functions include toroidal flux constraints, coil-length penalties, etc. 
Combined with analytical derivative calculations and optimization algorithms, the code optimizes coil configurations. 

\subsubsection{ FP3D code}

FP3D\cite{jiang2024fp3d} is an efficient numerical code for calculating 3D magnetic fields and particle motion in stellarators. It calculates 3D magnetic fields directly from coils or uses 3D fields from 
equilibrium codes such as VMEC or GVEC. The code can also use the NAE magnetic fields as input. 
The code has capabilities for magnetic-field-line tracing and test-particle simulations. In particular, the
neoclassical transport coefficients can be calculated. The drift-kinetic equation is used for computing particle motion.

\bibliographystyle{unsrt}
\bibliography{reference-paper8a}

\begin{thebibliography}{10}

\bibitem{post1987magnetic}
R.~F. Post.
\newblock The magnetic mirror approach to fusion.
\newblock {\em Nuclear Fusion}, 27(10):1579, 1987.

\bibitem{spitzer1958stellarator}
Lyman Spitzer.
\newblock The stellarator concept.
\newblock {\em Physics of Fluids}, 1(4):253, 1958.

\bibitem{azizov2012tokamaks}
E.~A. Azizov.
\newblock Tokamaks: from {A D Sakharov} to the present (the 60-year history of
  tokamaks).
\newblock {\em \mbox{Physics-Uspekhi}}, 55(2):190--203, 2012.

\bibitem{baldwin1977end}
D.~E. Baldwin.
\newblock End-loss processes from mirror machines.
\newblock {\em Reviews of Modern Physics}, 49(2):317--339, 1977.

\bibitem{rosenbluth1957stability}
M.~N. Rosenbluth and C.~L. Longmire.
\newblock Stability of plasmas confined by magnetic fields.
\newblock {\em Annals of Physics}, 1(2):120--140, 1957.

\bibitem{ivanov2017gas}
Alexander~Alexandrovich Ivanov and Vadim~Vadimovich Prikhodko.
\newblock Gas dynamic trap: experimental results and future prospects.
\newblock {\em \mbox{Physics-Uspekhi}}, 60(5):509, 2017.

\bibitem{feng2021proposal}
Zhichen Feng, Guodong Yu, Peiyou Jiang, and Guoyong Fu.
\newblock Proposal of a linked mirror configuration for magnetic confinement
  experiments.
\newblock {\em Nuclear Fusion}, 61(9):096021, 2021.

\bibitem{garren1991magnetic}
D.~A. Garren and A.~H. Boozer.
\newblock Magnetic field strength of toroidal plasma equilibria.
\newblock {\em Physics of Fluids B: Plasma Physics}, 3(10):2805--2821, 1991.

\bibitem{hindenlang2025computing}
Florian Hindenlang, Gabriel~G. Plunk, and Omar Maj.
\newblock Computing {MHD} equilibria of stellarators with a flexible coordinate
  frame.
\newblock {\em Plasma Physics and Controlled Fusion}, 67(4):045002, 2025.

\bibitem{zhu2018new}
Caoxiang Zhu, Stuart~R. Hudson, Yuntao Song, and Yuanxi Wan.
\newblock New method to design stellarator coils without the winding surface.
\newblock {\em Nuclear Fusion}, 58(1):016008, 2018.

\bibitem{jiang2024fp3d}
P.~Y. Jiang, Z.~C. Feng, G.~D. Yu, and G.~Y. Fu.
\newblock {FP3D}: A code for calculating {3D} magnetic field and particle
  motion.
\newblock {\em Physics of Plasmas}, 31(2):023906, 2024.

\bibitem{glowienka1981elmo}
J.~C. Glowienka.
\newblock {ELMO} bumpy torus: An alternate concept to tokamaks and mirrors.
\newblock {\em Journal of Vacuum Science and Technology}, 18(3):1088--1097,
  1981.

\bibitem{glagolev1981closed}
V.~M. Glagolev, B.~B. Kadomtsev, V.~D. Shafranov, and B.~A. Trubnikov.
\newblock Closed magnetic trap with rectilinear sections.
\newblock In {\em Proceedings of the 10th European Conference on Controlled
  Fusion and Plasma Physics}, volume~1, Moscow, 1981. European Physical
  Society.
\newblock Paper~\mbox{E-8}.

\bibitem{kondakov2000numerical}
V.~V. Kondakov, S.~F. Perelygin, and V.~M. Smirnov.
\newblock Numerical study of a {DRAKON}-type closed magnetic configuration with
  a spatial axis.
\newblock {\em Plasma Physics Reports}, 26(5):379--387, 2000.

\bibitem{mazzucato2020fusion}
E.~Mazzucato.
\newblock A fusion reactor scheme using magnetic mirrors.
\newblock {\em Fusion Science and Technology}, 76(7):807--813, 2020.

\bibitem{garren1991existence}
D.~A. Garren and A.~H. Boozer.
\newblock Existence of quasihelically symmetric stellarators.
\newblock {\em Physics of Fluids B}, 3(10):2822--2834, 1991.

\bibitem{kabel2003maxwell}
A.~C. Kabel.
\newblock {\mbox{Maxwell-Lorentz}} equations in general {Frenet-Serret}
  coordinates.
\newblock In {\em Proceedings of the 2003 Particle Accelerator Conference},
  volume~4, pages 2252--2254. IEEE, 2003.

\bibitem{boozer1981plasma}
A.~H. Boozer.
\newblock Plasma equilibrium with rational magnetic surfaces.
\newblock {\em Physics of Fluids}, 24(11):1999--2003, 1981.

\bibitem{rodriguez2023magnetohydrodynamic}
Eduardo Rodr{\'\i}guez.
\newblock Magnetohydrodynamic stability and the effects of shaping: a near-axis
  view for tokamaks and quasisymmetric stellarators.
\newblock {\em Journal of Plasma Physics}, 89(2):905890211, 2023.

\bibitem{shafranov1963equilibrium}
V.~D. Shafranov.
\newblock Equilibrium of a toroidal pinch in a magnetic field.
\newblock {\em Soviet Atomic Energy}, 13(6):1149--1158, 1963.

\bibitem{landreman2018direct}
Matt Landreman and Wrick Sengupta.
\newblock Direct construction of optimized stellarator shapes. {Part 1. Theory}
  in cylindrical coordinates.
\newblock {\em Journal of Plasma Physics}, 84(6):905840616, 2018.

\bibitem{landreman2019direct}
Matt Landreman, Wrick Sengupta, and Gabriel~G. Plunk.
\newblock Direct construction of optimized stellarator shapes. {Part 2.
  Numerical} quasisymmetric solutions.
\newblock {\em Journal of Plasma Physics}, 85(1):905850103, 2019.

\bibitem{plunk2019direct}
Gabriel~G. Plunk, Matt Landreman, and Per Helander.
\newblock Direct construction of optimized stellarator shapes. {Part 3.
  Omnigenity} near the magnetic axis.
\newblock {\em Journal of Plasma Physics}, 85(6):905850602, 2019.

\bibitem{landreman2020magnetic}
Matt Landreman and Rogerio Jorge.
\newblock Magnetic well and {Mercier} stability of stellarators near the
  magnetic axis.
\newblock {\em Journal of Plasma Physics}, 86(5):905860510, 2020.

\end{thebibliography}

\begin{samepage}
\par\medskip
\noindent\textbf{Acknowledgements}\par\nobreak
\begingroup
\small\normalfont
\noindent
We would like to thank Prof. Huichun Wu for useful discussions. We also thank Mr. Hengqian Liu for his assistance with the use of the FOCUS code. Finally, we thank Dr. Florian Hindenlang for the permission to use the GVEC code. 
\par
\endgroup
\end{samepage}

\end{document}